\documentstyle[12pt]{article}

\catcode`\@=11
\@addtoreset{equation}{section}

\newcommand{\be}{\begin{equation}}
\newcommand{\ee}{\end{equation}}
\newcommand{\ba}{\begin{eqnarray}}
\newcommand{\ea}{\end{eqnarray}}
\newcommand{\diag}{{\rm diag}}

\begin{document}

\baselineskip 24pt

\newcommand{\sheptitle}
{Non-Minimal Supersymmetric Higgs Bosons at LEP2}

\newcommand{\shepauthor}
{S. F. King$^1$ and P. L. White$^2$}

\newcommand{\shepaddress}
{$^1$Physics Department, University of Southampton, \\
Southampton, SO17 1BJ, UK. \\
Email {\tt king@soton.ac.uk} \\
and \\
$^2$Theoretical Physics, University of Oxford \\
1 Keble Road, Oxford OX1 3NP, UK. \\
Email {\tt plw@thphys.ox.ac.uk}
}

\newcommand{\shepabstract}
{We discuss the discovery reach of LEP2 for the Higgs sector of a
general extension of the MSSM including a single gauge singlet field.
This change introduces a new quartic Higgs boson self-coupling which
can increase the masses of the CP-even states, and also allows mixing
between singlet and non-singlet states which can reduce the couplings
of the mass eigenstates to the $Z$. The lightest CP-even Higgs boson is
bounded by a parameter $\Lambda$ which takes a maximum value
$\Lambda_{max}\approx 136-146$ GeV for top mass $150-195$ GeV. We
generalise the discussion  of the bound to include the entire CP-even
spectrum and show how  experiment may exclude values of $\Lambda$
smaller than some $\Lambda_{min}$. CP-even Higgs boson searches at LEP2
will be able to exclude $\Lambda_{min}\approx 81-105$  GeV, depending
on the machine parameters. We also present exclusion plots in the
$m_A-\tan\beta$ plane, based on an analysis of  CP-even, CP-odd and
charged Higgs production processes at LEP2.}

\begin{titlepage}
\begin{flushright}
SHEP 95-27, OUTP-95-31P\\
hep-ph/9508346
\end{flushright}
\vspace{.4in}
\begin{center}

{\large{\bf \sheptitle}}
\bigskip \\ \shepauthor \\ \mbox{} \\ {\it \shepaddress} \\ \vspace{.5in}
{\bf Abstract} \bigskip \end{center} \setcounter{page}{0}
\shepabstract

\end{titlepage}

\section{Introduction}
The minimal supersymmetric standard model (MSSM) \cite{revs} has a
Higgs sector consisting of two doublets $H_1$ and $H_2$ coupling to the
down-type quarks and charged leptons, and to the up-type quarks
respectively. The particle content, other than Goldstone bosons, is
then two CP-even states $h$, $H$; one CP-odd state $A$; and a charged
scalar $H^{\pm}$. The requirements of supersymmetry and gauge
invariance then constrain the quartic Higgs coupling in terms of the
gauge couplings, and we are left with only two free parameters to
describe the whole Higgs sector. These are conventionally taken to be
$m_A$, the mass of the CP-odd state, and $\tan\beta=\nu_2/\nu_1$ where
$\nu_i=\langle H_i^0\rangle$, and $\nu_1^2+\nu_2^2=\nu^2=(174{\rm
GeV})^2$. It is then straightforward to derive the formulae $m_h^2\le
M_Z$ and $m_{H^{\pm}}\ge M_W$ where the first of these in particular is
greatly affected by radiative corrections\cite{higgsRC}. Thus the
MSSM is quite constrained and leads to the usual LEP2 Higgs discovery
limits in the $m_A-\tan\beta$ plane.

If the Higgs sector of the MSSM is extended by the addition of further
particle content then much of this predictivity is lost. This occurs
for two reasons. Firstly, the constraint on the quartic couplings in
the Higgs sector is not purely a result of supersymmetry but also an
artefact of having only doublets in the Higgs sector, making it
impossible to introduce extra Yukawa couplings through the Higgs
superpotential in a way consistent with gauge invariance. Increasing
the particle content and including such Yukawa couplings destroys both
the upper bound on $m_h$ and the lower bound on $m_{H^{\pm}}$. Secondly, the
extra states which we can introduce will mix with the states present in
the MSSM, and can alter both their couplings and their masses.

For example, in the next-to-MSSM (NMSSM) where there is an extra gauge
singlet state $N$ which only couples through the Higgs superpotential,
and where there are only trilinear terms in the superpotential
\cite{NMSSM}, there are now three CP-even neutral Higgs scalars and two
CP-odd neutral Higgs scalars, due to the real and imaginary components
of the additional singlet scalar. In this model the lightest CP-even
neutral scalar may be significantly heavier than in the MSSM, and to
make matters worse this lightest CP-even scalar may have diluted
couplings to the Z boson due to the admixture of singlet component
\cite{NMSSM,dilution}. However, it is possible to derive a bound on the
lightest CP-even state, both in this and in more general models
\cite{NMSSMbound,boundRC,dilution}. Furthermore, in the limit that the
lightest CP-even Higgs boson is completely decoupled (and hence we
might think that the bound does not tell us anything about states which
could be detected), the bound applies instead to the second lightest
CP-even Higgs boson, while for light states which are merely weakly
coupled one can derive precise bounds on their heavier partners. These
become closer to the bound on the lightest as the singlet component of
the lightest becomes greater \cite{Kam}.

Our intention in this paper is to discuss how much of the parameter
space of such a general SUSY model can be covered by searches at LEP2,
comparing them closely with the corresponding results for the MSSM. We
shall consider the most general possible model with a gauge singlet and
show that LEP2 can discover CP-even Higgs states in any part of the
$m_A-\tan\beta$ plane, and can exclude significant parts of this plane
even with the most general possible mixing.

We begin with a short review of the MSSM and its simplest extensions
and how they affect the mass matrices. In section 4, we discuss the
conventional bound on the lightest CP-even state in this model. In
section 5, a number of search strategies are discussed, and we consider
how effective the various MSSM searches are in an extended model.
Finally, in section 6 we present results applicable to LEP2. Section 7
is the conclusion.

\section{The MSSM}
Here the superpotential is:
\begin{equation}
W_{MSSM}=-\mu H_1H_2+\ldots
\end{equation}
where $H_1H_2=H_1^0H_2^0-H_1^-H_2^+$.

The superpotential leads to the tree-level Higgs potential:
\ba
V_{MSSM} & = & m_{1}^2|H_1|^2 +m_{2}^2|H_2|^2 +
m_{12}^2(H_1H_2+H.c.) \nonumber \\
         & + & \frac{1}{8}(g^2+g'^2)(|H_1|^2-|H_2|^2)+
\frac{g^2}{2}|H_1^{\ast}H_2|^2
\ea
where $m_{1}^2=m_{H_1}^2+\mu^2$, $m_{2}^2=m_{H_2}^2+\mu^2$,
and $m_{12}^2=-B\mu$ where $B$ and $m_{H_i}^2$ are the usual soft
SUSY-breaking terms.

The tree-level CP-odd mass-squared matrix is
\begin{equation}
M_{A}^2=
\left(\begin{array}{cc}
m_{12}^2t_{\beta}& m_{12}^2  \\
m_{12}^2 & m_{12}^2/t_{\beta}
\end{array} \right)
\label{CP-odd}
\end{equation}
$M_A^2$ is diagonalised by
$UM_{A}^2U^{\dagger}=\diag(m_A^2,0)$ where $U$ is given by
\begin{equation}
U=\left(\begin{array}{cc}
     s_{\beta} & c_{\beta}  \\
     -c_{\beta} & s_{\beta}
   \end{array} \right)
\label{U}
\end{equation}
and $m_A$ is given by $m_A^2=m_{12}^2/(s_{\beta}c_{\beta})$ where
$s_{\beta}=\sin\beta$, $c_{\beta}=\cos\beta$ and $t_{\beta}=\tan\beta$.

The tree-level charged Higgs mass squared matrix is
\begin{equation}
M_{H^{\pm}}^2=
\left(\begin{array}{cc}
M_{W}^2s_{\beta}^2 + m_{12}^2t_{\beta}
& M_{W}^2s_{\beta}c_{\beta} + m_{12}^2  \\
M_{W}^2s_{\beta}c_{\beta} + m_{12}^2
& M_{W}^2c_{\beta}^2 + m_{12}^2/t_{\beta}
\end{array} \right)
\label{Hpm}
\end{equation}
$M_{H^{\pm}}^2$ is diagonalised by
$UM_{H^{\pm}}^2U^{\dagger}=\diag(m_{H^{\pm}}^2,0)$ where
$U$ is as before and
$m_{H^{\pm}}$ is given by
$m_{H^{\pm}}^2=M_W^2+m_A^2$.

The tree-level CP-even Higgs mass-squared matrix is,
after substituting the tree-level expression for $m_{12}^2$
in terms of $m_A^2$,
\begin{equation}
M^2=
\left(\begin{array}{cc}
M_{Z}^2c_{\beta}^2 + m_{A}^2s_{\beta}^2
& -(M_{Z}^2+m_{A}^2)s_{\beta}c_{\beta}   \\
- (M_{Z}^2+m_A^2)s_{\beta}c_{\beta}
& M_{Z}^2s_{\beta}^2 + m_{A}^2c_{\beta}^2
\end{array} \right)
\label{CP-even}
\end{equation}
$M^2$ is diagonalised by
$VM^2V^{\dagger}=\diag(m_{H}^2,m_{h}^2)$ where
\begin{equation}
V=
\left(\begin{array}{cc}
c_{\alpha} & s_{\alpha}  \\
-s_{\alpha} & c_{\alpha}
\end{array} \right)
\label{V}
\end{equation}
where $-\pi/2\leq \alpha \leq 0$ and $m_{h}^2,m_{H}^2$
are given by
\begin{equation}
m_{H,h}^2=\frac{1}{2}(m_A^2+M_Z^2)
\pm \frac{1}{2} \sqrt{(m_A^2+M_Z^2)^2-4m_A^2M_Z^2c_{2\beta}^2}
\label{hmass}
\end{equation}
and the two angles are related by:
\begin{equation}
\tan 2\alpha =\tan 2\beta \frac{(m_A^2+M_Z^2)}{(m_A^2-M_Z^2)}
\end{equation}
In the above conventions, the ZZh coupling has an additional
factor of $\sin (\beta -\alpha)$ relative to the standard model,
and the ZZH coupling has a factor of $\cos (\beta - \alpha )$.
The ZhA coupling is proportional to $\cos (\beta - \alpha )$.

It is interesting to observe that if we do not diagonalise $M^2$ but
instead rotate to a primed basis where the second Higgs doublet does
not have any VEV, then the 11 element of the matrix no longer contains
any $m_A$ dependence. Since the smallest eigenvalue of a real symmetric
matrix is less than or equal to the smallest diagonal term, this gives
a useful upper bound on the lightest CP-even Higgs mass. This rotation
is given by $WM^2W^{\dagger}=M'^2$ where
\begin{equation}
W=
\left(\begin{array}{cc}
c_{\beta} & s_{\beta}  \\
-s_{\beta} & c_{\beta}
\end{array} \right)
\label{W}
\end{equation}
and the bound is then given by
\be
m_h^2\leq M^{\prime 2}_{11}=M_Z^2\cos^22\beta.
\ee
Starting from the primed basis
the CP-even Higgs matrix is diagonalised by
\be
XM'^2X^{\dagger}=\diag(m_{H}^2,m_{h}^2)
\ee
where $X=VW^{\dagger}$.

Note that in this basis the additional factor in the ZZh coupling
relative to the standard model of $\sin(\beta-\alpha)$ is simply
$X_{21}$, and similarly the ZZH factor of $\cos(\beta-\alpha)$
is just $X_{11}$.

\section{The MSSM plus a Singlet}
Now let us consider a very general Higgs superpotential of the form
\begin{equation}
W=W_{MSSM}+\lambda NH_1H_2+f(N)
\end{equation}
where $f$ is an arbitrary holomorphic function of $N$, the singlet
field, and $H_1$ and $H_2$ are the usual Higgs doublets coupling to
down and up quarks respectively. Since the singlet does
not have gauge couplings we may then write
\ba
V &=&V_{MSSM}+\lambda^2 \bigl |H_1H_2 \bigr |^2 \cr
          &&  +\Bigl(\lambda (N+\bar N)\mu +\lambda^2|N|^2 \Bigr)
                               \Bigl ( |H_1|^2+|H_2|^2 \Bigr ) \cr
          &&  +\Bigl ( \lambda \frac{\partial \bar f}{\partial \bar N}
                    - \lambda N A_\lambda \Bigr ) H_1H_2 +h.c.
             +\cdots
\ea
where the ellipsis indicates terms with no dependence on $H_1$ or
$H_2$. Given that $V_{MSSM}$ includes terms $\mu^2(|H_1|^2+|H_2|^2)$
and $B\mu H_1H_2$ it is then clear that we can account for the
effects of all the singlet dependent terms on the upper 2$\times$2
block of the mass matrices (CP-odd, CP-even, and charged) purely by
redefinitions of $B\rightarrow B'$ and
$\mu\rightarrow \mu'$ in terms of the VEV of the singlet
$<N>=x$ and
by including the effects of the $\lambda$ dependent quartic term.

We further note that we could obtain identical results with an arbitrary
number of singlets, since we can always rotate so that only one of them
couples to $H_1H_2$ and the remainder simply complicate the form of $f$
and of the soft terms, which we are essentially regarding as arbitrary.
With one singlet, the most general superpotential is
\begin{equation}
W =-\mu H_1H_2 + \lambda NH_1H_2 - \frac{k}{3}N^3
  + \frac{1}{2}\mu^{\prime}N^2 + \mu^{\prime\prime}N
\label{Wgeneral}
\end{equation}
In the limit that $\mu,\mu^{\prime}\mu^{\prime\prime}\rightarrow 0$, the
above superpotential reduces to that of the NMSSM\cite{NMSSM}, while if
$N$ is removed it reduces to that of the MSSM.

The mass matrix for the CP-odd scalars
in the basis $(H_1,H_2,N)$ is a complicated
3-dimensional generalisation of $M_A^2$ in the MSSM.
With only minimal risk of ambiguity we shall use the same notation
for such generalisations as for the MSSM.
The tree-level CP-odd mass squared matrix becomes :
\begin{equation}
M_{A}^2=
\left(\begin{array}{ccc}
m_{12}'^2t_{\beta}& m_{12}'^2 & . \\
m_{12}'^2 & m_{12}'^2/t_{\beta} & . \\
. & . & .
\end{array} \right)
\label{CP-odd2}
\end{equation}
where $m_{12}'^2= -B'\mu'$.
The CP-odd matrix has its Goldstone modes isolated by
\begin{equation}
UM_{A}^2U^{\dagger} \equiv
\left(\begin{array}{ccc}
0 & 0 & 0 \\
0 & m_A'^2 & . \\
0 & . & .
\end{array} \right)
\end{equation}
where $U$ is given by
\begin{equation}
U=
\left(\begin{array}{ccc}
c_{\beta} & -s_{\beta} & 0 \\
s_{\beta} & c_{\beta} & 0 \\
0 & 0 & 1
\end{array} \right)
\label{U2}
\end{equation}
Clearly $m_A'$ is the analogue of $m_A$ of the MSSM,  and
$m_A'^2=m_{12}'^2/(s_{\beta}c_{\beta})$. The entries in these matrices
represented by dots are complicated functions of soft terms and
parameters from $f$ and the extended soft potential, and since $f$ is
arbitrary they are essentially unconstrained.  We define the matrix
$V_A$ which diagonalises $M_A^2$ as follows
\be
(V_AU)M_{A}^2(V_AU)^{\dagger} = \diag(0,m_{A_1}^2,m_{A_2}^2)
\ee
where we have taken $m_{A_1}<m_{A_2}$ and
$V_A$ can be represented in terms of one mixing angle $\gamma$
\begin{equation}
V_A=
\left(\begin{array}{ccc}
1 & 0 & 0 \\
0 & c_{\gamma} & s_{\gamma} \\
0 & -s_{\gamma} & c_{\gamma}
\end{array} \right)
\label{gamdef}
\end{equation}

As regards the charged Higgs sector,
the singlets obviously cannot mix with charged scalars, and we find
that (at tree-level) the mass of the charged Higgs is given by
\be
m_{H^{\pm}}^2=m_A'^2+M_W^2-\lambda^2\nu^2
\ee
We can immediately obtain the MSSM results by simply setting
$\lambda=0$ and removing all of the singlet terms from the mass
matrices. Clearly a non-zero $\lambda$ tends to reduce the
charged scalar masses which can be arbitrarily small.

The CP-even mass squared matrix  is now a 3$\times$3 matrix in the
basis $(H_1,H_2,N)$ which may be written as
\begin{equation}
M^2=
\left(\begin{array}{ccc}
M_Z^2c_{\beta}^2+m_A'^2s_{\beta}^2 &
-(M_Z^2+m_A'^2-2\lambda^2\nu^2)s_{\beta}c_{\beta}   & .   \\
-(M_Z^2+m_A'^2-2\lambda^2\nu^2)s_{\beta}c_{\beta}     &
M_Z^2s_{\beta}^2+m_A'^2c_{\beta}^2 & . \\
. & . & .
\end{array} \right)
\label{CPevenmass}
\end{equation}
where, as before, the dots indicate complicated entries.
As in the MSSM the CP-even Higgs matrix is diagonalised by
\be
VM^2V^{\dagger}=\diag(m_{h_1}^2,m_{h_2}^2,m_{h_3}^2)
\label{diag3}
\ee
where
$V$ is some complicated 3$\times$3 matrix.
Note that we have now chosen to order the
mass eigenstates as
\be
m_{h_1}<m_{h_2}<m_{h_3}
\label{ordering}
\ee
so that the definition of $V$ here
does {\em not} reduce to that of $V$ defined
earlier where the eigenvalues are conventionally ordered oppositely.
For example in the MSSM if we had required that the
CP-even Higgs matrix were diagonalised as
$VM^2V^{\dagger}=\diag(m_{h}^2,m_{H}^2)$
then $V$ would have to include a re-ordering of the mass eigenstates,
and would have been given by
\begin{equation}
V =
\left(\begin{array}{cc}
-s_{\alpha} & c_{\alpha}  \\
-c_{\alpha} & -s_{\alpha}
\end{array} \right)
\end{equation}
where $\alpha$ is the angle defined earlier in the MSSM.
It is to this form that our generalised $V$ must
reduce in the MSSM limit.

As in the MSSM we observe that if we do not diagonalise
$M^2$ but instead rotate to a basis where the second
Higgs doublet does not have any VEV, then the 11 element
of the matrix gives a useful upper bound on the lightest
CP-even Higgs mass \cite{NMSSMbound}. This rotation is given by
$WM^2W^{\dagger}=M'^2$ where
\begin{equation}
W=
\left(\begin{array}{ccc}
c_{\beta} & s_{\beta} & 0 \\
-s_{\beta} & c_{\beta} & 0 \\
0 & 0 & 1
\end{array} \right)
\label{W2}
\end{equation}
and the bound is given by
\be
m_{h_1}^2 \leq M'^2_{11}=
M_Z^2\cos^2 2\beta + \lambda^2\nu^2\sin^22\beta
\label{bound1}
\ee
The reason why this bound is useful is simply that
it has no ${m'_A}^2$ dependence.
For example a less useful bound is,
\be
m_{h_1}^2 \leq M'^2_{22}=
(M_Z^2- \lambda^2\nu^2)\sin^22\beta + {m'_A}^2
\label{bound1lessuseful}
\ee

Starting from the primed basis
the CP-even Higgs matrix is diagonalised by
\be
XM'^2X^{\dagger}=\diag(m_{h_1}^2,m_{h_2}^2,m_{h_3}^2)
\label{XX}
\ee
where $X=VW^{\dagger}$.

We shall define the relative couplings
$R_i\equiv R_{ZZh_i}$ as the $ZZh_i$ coupling in units of the standard model
$ZZh$ coupling, and similarly we shall define a $Zh_iA_j$ coupling
factor $R_{Zh_iA_j}$.
For example $R_{ZZh_1}$ is a generalisation of $\sin (\beta - \alpha)$
and the $R_{Zh_1A_i}$ are generalisations of $\cos (\beta - \alpha)$
in the MSSM. In our notation we find, using the results of
Ellis {\em et al} \cite{NMSSM},
\ba
R_{i} & = & \cos \beta V_{i1} + \sin \beta V_{i2} \nonumber \\
                    & = & (VW^{\dagger})_{i1}=X_{i1}.
\label{ZZh}
\ea
The $Zh_iA_j$ coupling factorises
into a CP-even factor $S_i$ and a CP-odd factor $P_j$\cite{NMSSM}
\be
\label{ZhA}
R_{Zh_iA_j}=S_i P_{j}
\ee
where
\ba
S_i & = & -V_{i1}\sin \beta + V_{i2}\cos \beta  \nonumber \\
            & = & (WV^{\dagger})_{2i} = X_{i2}
\label{CP-evenfactor}
\ea
and
\be
P_j = (V_AU)_{(j+1)2}\cos \beta +(V_AU)_{(j+1)1}\sin \beta
\label{CP-oddfactor}
\ee
Eq.(\ref{CP-oddfactor}) implies the simple intuitive results
\be
P_1=\cos \gamma, \ \ P_2=-\sin \gamma
\label{P}
\ee
where $\gamma$ defined in Eq.(\ref{gamdef})
is the angle which controls the amount of singlet
mixing in the CP-odd sector.

\section{Implementing the Bound on $h_1$}

In this section we shall consider the absolute upper bound
on the mass of the lightest CP-even state in this model,
using Eq.(\ref{bound1}) plus radiative corrections.
For notational ease we shall first define
\be
\Lambda^2 \equiv M'^2_{11}.
\label{Lambda}
\ee
Thus the bound is simply
\be
m_{h_1}^2 \leq \Lambda^2
\ee
Clearly $\Lambda^2$ is a function of $\tan \beta$ and $\lambda$,
$\Lambda^2(\tan \beta,\lambda)$,
and to find the absolute upper bound we must maximise this
function so that
\be
m_{h_1}^2 \leq \Lambda_{max}^2
\ee
where $\Lambda_{max}^2$ is the maximum value of
$\Lambda^2(\tan \beta$ and $\lambda)$,
\be
\Lambda_{max}^2 \equiv max(\Lambda^2(\tan \beta,\lambda)).
\ee
Our task in this section is therefore to find
$\Lambda_{max}^2$ in the presence of radiative corrections.

It is well known that radiative corrections, which we have so far
ignored, drastically affect the bound \cite{boundRC}.
In order to deal with these radiative corrections many techniques have
been proposed \cite{higgsRC}. Here we shall follow the method proposed
in ref.\cite{ceqw}, which we shall briefly review. According to this
method a scale $M_{susy}$ is defined by  $M_{susy}^2=(m_{\tilde
t_1}^2+m_{\tilde t_2}^2)/2$ where $m_{\tilde t_i}$ are the stop mass
eigenvalues, and the couplings $h_t$ and $h_b$ found at this scale.
Here the squarks are integrated out, leaving an effective Higgs
potential involving Yukawa couplings with boundary conditions which
differ from the tree-level ones by some finite corrections
\cite{higgsRC}. Then the Yukawa couplings (including $h_t$) are run
from $M_{susy}$ down to $m_t$. Below $m_t$ the Yukawa couplings are
approximately constant, and so the Higgs potential may be minimised at
this scale. The effects of the RG running have been estimated
analytically, leading to fully analytic results for the Higgs masses
which agree well with more elaborate methods \cite{ceqw}.

For $\lambda=0$ (as in the MSSM) the bound is obviously largest for
$\cos^2(2\beta)=1$. Given the restricted range $\pi/4 \leq \beta \leq
\pi/2$ this implies that the bound is maximised for $\beta = \pi/2$,
$\cos 2\beta =-1$, corresponding to $\tan \beta = \infty$ (or in
practice its maximum allowed value). However, for sufficiently large
$\lambda$ the tree-level bound will be maximised for $\tan\beta$ equal
to its minimum value. In order to obtain an absolute upper bound, we
must thus derive an upper limit on $\lambda$, which will turn out to be
a function of $\tan\beta$. This can be done by demanding that all of
the Yukawa couplings remain perturbative up to the GUT scale
\cite{NMSSMbound,boundRC,dilution}.

We have used the newly discovered mass of the top quark \cite{CDF} and
the two loop SUSY RG equations in the NMSSM \cite{epic} to derive an
upper bound on $\lambda$ (defined at $m_t$, and calculated by requiring
all Yukawa couplings $h_t$, $h_b$ and $\lambda$ to remain perturbative
up to a scale of $10^{16}$GeV), which we show as a function of
$\tan\beta$ in Figure 1. Results are shown for various values of $m_t$
and $\alpha_3(M_Z)$, and it is likely that the errors from these two
measurements will greatly dominate any other uncertainties of our
calculation. The main features of this graph are that $\lambda$ has an
upper bound of around 0.5 to 0.7 for intermediate $\tan\beta$, while
for large (small) $\tan\beta$ the triviality constraints on $h_b$
($h_t$) force this upper bound very rapidly to zero.

In Figure 1, we have assumed that the effect of any other Yukawa
couplings in the singlet sector of the superpotential is negligible,
but in fact such couplings can have a significant impact, as shown in
Figure 2, where we fix $\tan\beta$ and explore the dependence of the
upper bound on $\lambda$ as a function of the coupling $k$ defined in
Eq.~(\ref{Wgeneral}) for a model with only one singlet. The most
important feature of this figure is that it clearly displays how adding
extra Yukawa couplings will always {\em reduce} the bound on $\lambda$,
since these will always appear with the same sign in the RG equations.
The reason why $\lambda$ falls so rapidly to zero above a certain value
of $k$ is that here $k$ is approaching its triviality limit, while for
small $k$ its impact on the upper bound of $\lambda$ is negligible.

In Figures 3a to 3c, we plot the bound $\Lambda$ as a function of
$\tan\beta$, $M_{susy}$, and $m_t$ respectively. For each of these
figures, we set the squark soft masses squared $m_Q^2$, $m_T^2$, and
$m_B^2$ to be equal, and we allow the trilinear squark--Higgs coupling
$A_t$, $\lambda$ and, except in Figure 3a, $\tan\beta$ to take on the
values which maximise the bound. In Figures 3a,3c we select $M_{susy}$
as defined previously to be 1TeV, and in Figures 3a and 3b we select
$m_t$=175GeV. We display bounds for both the MSSM (lower dashed line)
and its extension with a singlet. Note that for extreme values of
$\tan\beta$ the maximum allowed value of $\lambda$ falls rapidly to
zero, and so the two bounds are then the same, while the same happens
for very large $m_t$ since here $h_t$ is again close to triviality.
Thus the large value of $m_t$ preferred by $CDF$ \cite{CDF} means that
the two bounds are now within around 8GeV. However, it is important to
realise that one of the main problems for LEP2 will be that for the
MSSM small $\tan\beta$ always gives a light CP-even Higgs, while this
need no longer be true in a non-minimal extension, thus seriously
damaging the prospects for particle searches.

\section{LEP2 Search Strategies}

\subsection{$Z\rightarrow Zh_i$ and General Bounds on CP-even Higgses}

Recently upper bounds on all three neutral CP-even Higgs scalars in the
general extension of the MSSM with a single gauge singlet superfield
have been derived \cite{Kam}. The basic observation which follows from
Eq.(\ref{XX}) and the definition $M'^2_{11}\equiv \Lambda^2$ is:
\footnote{Note that $X$ is real if the Higgs sector conserves CP.}
\ba
\Lambda^2&=& X_{11}^2m_{h_1}^2+
X_{21}^2m_{h_2}^2+ X_{31}^2m_{h_3}^2 \cr
 &=&R_1^2m_{h_1}^2+R_2^2m_{h_2}^2+R_3^2m_{h_3}^2
\label{basic}
\ea
where we have used the fact that $R_i=X_{i1}$.
Eq.(\ref{basic}) together with Eq.(\ref{ordering})
clearly implies
\be
\Lambda^2\geq (R_1^2+R_2^2+R_3^2)m_{h_1}^2 \nonumber
\ee
and given that
\be
R_1^2+R_2^2+R_3^2=1
\label{unitarity}
\ee
by the unitarity of $X$,
we find the usual bound in Eq.(\ref{bound1}),
\be
m_{h_1}^2\leq \Lambda^2
\ee
Using a similar argument,
a bound on the $h_2$ mass may be obtained from
Eq.(\ref{basic}) :
\be
m_{h_2}^2 \leq \frac{\Lambda^2-R_1^2m_{h_1}^2}{1-R_1^2}
\label{bound2}
\ee
$m_{h_3}^2$ is given from Eq.(\ref{basic}) by
\begin{equation}
m_{h_3}^2 =
\frac{\Lambda^2-R_1^2m_{h_1}^2-R_2^2m_{h_2}^2}
{1-R_1^2-R_2^2}
\label{equality3}
\end{equation}
from which we can extract a bound
\be
m_{h_3}^2 \leq
\frac{\Lambda^2-(R_1^2+R_2^2)m_{h_1}^2}{1-R_1^2-R_2^2}
\label{bound3}
\ee
As we shall see, Eqs.(\ref{bound2}) and (\ref{bound3}) are useful upper
bounds on $m_{h_2}$ and $m_{h_3}$ in terms of $m_{h_1}^2$, $R_1$, $R_2$
and $\Lambda^2\equiv M'^2_{11}$ (which is just the bound on
$m_{h_1}^2$). These bounds tightly constrain the spectrum in terms of
the couplings and hence will allow us to study  the reach of colliders
when the mixing parameters take on arbitrary values.

The above general bounds are useful
since $h_1$ may be light but very weakly
coupled. In the MSSM this can happen when $\sin (\beta - \alpha)$
becomes small. In the present model $R_1$ can also be small because
$h_1$ may contain a large singlet component. In this case
it does not matter how light $h_1$ is since its couplings may be
so weak that it can never be detected. At first sight this seems to be
a catastrophe for LEP2 where in the MSSM $h_1$ is usually the easiest
state to find. However, if $R_1\approx 0$ we may simply ignore $h_1$
and concentrate on $h_2$ which then becomes the lightest physically
coupled CP-even state. Moreover it can be seen from Eq.(\ref{bound2})
that if $R_1\rightarrow 0$ then $h_2$ must satisfy
\be
m_{h_2}^2 \leq \Lambda^2
\label{bound22}
\ee
which is just the bound on $m_{h_1}$. In other words, if the lightest
CP-even state is essentially invisible, then the second lightest CP-even
state must be quite light, and may be within the reach of LEP2.
Similarly if both $R_1$ and $R_2$ are $\approx 0$ then Eq.(\ref{equality3})
implies that
\be m_{h_3}^2=\Lambda^2
\label{bound33}
\ee
The above restrictions on the Higgs masses
and couplings imply that a collider of a
given energy and integrated luminosity will be able to exclude values
of $\Lambda$ below which
it is impossible for all the Higgses to have simultaneously
escaped detection.

We shall discuss this further in Section 5.3.

\subsection{Exclusion Plots in the $R^2-m_h$ Plane at LEP}

In this sub-section we  we shall be concerned with exclusion plots  in
the $R^2-m_h$ plane such as those that have already been obtained at
LEP1 and whose possible form at LEP2 has been predicted, where
$R^2\equiv \sin^2(\beta - \alpha)$ in the MSSM. LEP1 has not discovered
a CP-even Higgs boson \cite{LEPex}, and this non-observation has
enabled a $95\%$ CL exclusion limit to be extracted on the value of the
square of the relative $ZZh$ coupling $R^2$, as a function of the
CP-even Higgs mass $m_h$ extending out to $m_h=65$ GeV\cite{mRLEP1}.
Similar exclusion plots at LEP2 will extend the Higgs mass range to
$m_h>65$ GeV leading to possible exclusion curves whose precise shape
will depend on the energy and integrated luminosity of
LEP2\cite{janot}. Three different sets of LEP2 machine parameters have
been considered corresponding to energies and integrated luminosities
per experiment of \cite{janot}
\ba
\sqrt{s} & = & 175 \ {\rm GeV},\ \ \int{\cal{L}}=150
 \ {\rm pb}^{-1} \nonumber \\
\sqrt{s} & = & 192 \ {\rm GeV},\ \ \int{\cal{L}}=150
 \ {\rm pb}^{-1} \nonumber \\
\sqrt{s} & = & 205 \ {\rm GeV},\ \ \int{\cal{L}}=300
 \ {\rm pb}^{-1} .
\label{LEP2}
\ea
The existing LEP1 \cite{mRLEP1} and anticipated LEP2 \cite{janot}
exclusion plots are combined in Fig.4a, for the three different LEP2
scenarios, and  assuming a 100\% $h_1$ branching fraction into
$\bar{b}b$. The very steep rise of the contours means that, if a
standard model-like Higgs boson of a certain fixed mass can be produced
at LEP (corresponding to $R^2=1$) then reducing $R_{ZZh}^2$ has little
effect on its visibility until $R_{ZZh}^2$ becomes quite small. The
relatively flat LEP1 regions of the contours correspond to minimum
threshold values of $R_{ZZh}^2$ below which nothing can be seen for any
value of the mass.

The above exclusion plots in Fig.4a are the result of a sophisticated
Monte Carlo simulation of the LEP detectors. It is interesting to
observe that these contours roughly correspond to the value of the
coupling $R^2$ which would yield 50 Higgs events of a given mass. The
plot of contours of 50 $h$ events as a function of $m_{h}$ and $R^2$
for the three different LEP2 machine scenarios is shown for comparison
in Fig.4b. Note that this simple parameterisation of the exclusion
plots of Fig.4a fails badly for $m_h \approx M_Z$, and also the actual
exclusion limits from LEP1 are significantly stronger than our LEP2
approximation for small Higgs masses. Nevertheless Fig.4b provides a
useful caricature of the exclusion plots in Fig.4a.

Finally note that the exclusion plots in Fig.4 may be interpreted for
each of the three CP-even Higgs bosons in this model taken separately.
In other words the disallowed region to the  upper and left of the
lines will exclude values of ${R_1}^2$ for a given $m_{h_1}$, and
similarly for the other two CP-even Higgs bosons.

\subsection{Excluded Values of $\Lambda$}

Although there are  a large number of parameters in this model, and
three CP-even Higgs bosons, it proves possible to exclude values of
$\Lambda$ smaller than a certain amount, where $\Lambda$ is defined as
above to be the upper bound on the lightest CP-even Higgs boson,  and
is to be regarded as a function of $\lambda, \tan \beta$ and all the
other parameters which enter the calculation of radiative corrections
such as the top mass, the squark masses, and so on. If LEP2 does not
discover any CP-even Higgs bosons then exclusion plots such as those in
Fig.4a may be produced. We now show that  such exclusion plots may be
used to place an excluded lower limit on the value of $\Lambda$ in this
model. If the excluded lower limit on $\Lambda$ reaches the theoretical
upper limit of about 146 GeV (dependent on the top mass and SUSY
spectrum as discussed above) then the model is excluded. For example, a
linear collider of energy 300 GeV should be able to exclude the
model \cite{Kam}.  We now describe how LEP may be used to exclude values
of $\Lambda$ in this model.

Clearly a specified value of $\Lambda$ is consistent with many sets of
values of the parameters $m_{h_i}$, $R_i$ subject to the bounds
discussed in section 5.1. In general some of these sets of $m_{h_i}$,
$R_i$ will lead to one or more CP-even Higgs boson in the excluded
(upper left) part of Fig.4a and some sets will lead to Higgs bosons
only in the allowed region. This means that, for a given $\Lambda$, one
or more CP-even Higgs bosons may or may not be discovered at LEP2,
depending on the  values of the other (complicated unknown) parameters
in the model. According to our discussion in Section 5.1, it is clear
that as $\Lambda$ is reduced, more and more of the allowed sets of
$m_{h_i}$, $R_i$ will move into the excluded part of Fig.4a. As
$\Lambda$ is reduced below some critical value all the sets of values
of $m_{h_i},R_i$ will eventually fall into the excluded region of
Fig.4a. This critical value of $\Lambda$ is the maximum value of
$\Lambda$ excluded by LEP. This implies that the maximum excluded value
of $\Lambda$ is determined by  whether the ``worst case'' ({\em i.e.}
hardest to see experimentally) values of $m_{h_i},R_i$ consistent with
this value of $\Lambda$ lie within the allowed region or not.

In order to determine the ``worst case'' parameters we first fix
$\Lambda$  at some specified value (less than its maximum as determined
in section 4) and then we scan over values of $m_{h_1}$ from zero up to
$\Lambda$. For each value of $m_{h_1}$ we set $R_1$ equal to the
maximum allowed value as shown in Fig.4a (or 1 if this would require
$R_1>1$). For $\Lambda$, $m_{h_1}$ and $R_1$ fixed as above, we then
scan over $m_{h_2}$ from $m_{h_1}$ up to its upper bound which is now
fixed according to Eq.(\ref{bound2}). For each value of $m_{h_2}$ we
set $R_2$ equal to its maximum allowed value as before.  For $\Lambda$,
$m_{h_1}$, $m_{h_2}$, $R_1$ and $R_2$ fixed as above, $m_{h_3}$ and
$R_3$ are now completely specified by Eqs.(\ref{unitarity}) and
(\ref{equality3}). If $R_3$ is larger than its maximum allowed value
(according to Fig.4b), then the ``worst case'' with the given values of
$m_{h_1}$ and $m_{h_2}$ is excluded. If all the ``worst case'' points
in the $m_{h_1}$, $m_{h_2}$ scan turn out to be excluded then we
conclude that this value of $\Lambda$ is excluded by LEP.

According to Eq.(\ref{bound1}), the value of $\Lambda^2\equiv
{M'_{11}}^2$ is a function of $\lambda$ and $\tan \beta$, plus
parameters which enter in the radiative corrections. For a given value
of $\lambda$, and given radiative correction parameters, the exclusion
limit on $\Lambda$ may be interpreted as an exclusion limit on $\tan
\beta$ in this model, independently of $m'_A$. Thus the present
analysis will yield a horizontal exclusion line in the $m'_A-\tan
\beta$ plane, as we shall see in Section 6.

If the observability criterion above is replaced by the
approximate exclusion limit obtained from
50 higgs events regardless of branching
fractions, as in Fig.4b, then the above algorithm can be solved
analytically. In this simple case,
the ``worst case'' for LEP2 to discover would be one
where the largest of the three cross-sections $\sigma_i$ was minimised,
where
\be
\sigma_i= \sigma (e^+e^- \rightarrow Zh_i)=
\sigma_{SM} (e^+e^- \rightarrow Zh)|_{m_h=m_{h_i}}R_i^2
\ee
It is not hard to show that this
will occur when $\sigma_1=\sigma_2=\sigma_3$, and hence when
$\sum_i\sigma_{SM}(m_{h_i})R_i^2$
is minimised.
Using the constraints in Eqs.(\ref{basic}) and (\ref{unitarity}),
together with the analytical form for the
mass dependence of the tree-level cross-section,
\be
\sigma_{SM}(m_{h_i})\sim \lambda^{1/2}(\lambda+12sm_Z^2)
\ee
where $\lambda=(s+(M_Z+m_h)^2)(s-(M_Z-m_h)^2)$, it is then
straightforward to prove that this will always be minimised when all
three masses are degenerate and $R_i^2=1/3$.

This simple result leads to the contour plot of
$\Lambda=\sqrt{M^{\prime 2}_{11}}$ in the integrated luminosity-energy
plane shown in Fig.5. \footnote{Note that this is {\it total}
luminosity, not luminosity per experiment as elsewhere in this paper.}
Along each of the contours the energy and integrated
luminosity correspond to at least one CP-even Higgs boson being
produced with a yield of 50 events, for any allowed choice of $m_{h_1}$
and $m_{h_2}$. Thus this contour plot is an estimate of the values of
$\Lambda$ which may be excluded for different machine parameters. For
example, LEP2 with an energy of 205 GeV will place a limit
$\Lambda>100$ GeV, depending on the integrated luminosity. For this
energy there is a rapid increase in the reach of $\Lambda$ from 10-100
GeV as the luminosity is increased from 100-400 pb${}^{-1}$, followed
by a very slow increase if the luminosity is increased beyond this.

To summarise, the worst case parameters are when all the three CP-even
Higgs bosons have a mass equal to $m_{h_i}=\Lambda$ and equally
suppressed couplings ${R_i}^2=1/3$. In fact since in this case there
are three Higgs bosons to discover the cross-section will be three
times larger than for each Higgs boson taken separately. Nevertheless,
it is clear that for sets of parameters when the Higgs boson masses
are not degenerate and the Higgs bosons must be considered separately
the worst case will never be worse than that just described. The
excluded value of $\Lambda$ is therefore simply determined by the
following rule of thumb: consider a single Higgs boson with a coupling
equal to $R^2=1/3$, and find its maximum excluded mass for a given set
of machine parameters, then equate this mass with the maximun excluded
value of $\Lambda$.

According to this rule of thumb, LEP1 already places a limit on
$\Lambda$ of $\Lambda>59$ GeV, which is just equal to the mass limit
for a CP-even Higgs boson with its ZZh coupling suppressed by
$R^2=1/3$. Note that because of the steep rises of the function plotted
in Fig.4a, the excluded values of $\Lambda$ are not far from the
present excluded value of SM Higgs boson mass of about 65 GeV.
Similarly we find that LEP2 will yield the exclusion limits
$\Lambda>\Lambda_{min}=81,93,105$ GeV, for the three sets of LEP2
machine parameters in Eq.(\ref{LEP2}), respectively.

The above rule of thumb is easily extended to the more general case of
$n$ singlets being added to the MSSM, {\em i.e.} the (M+$n$)SSM. In such
a model the worst case will correspond to $2+n$ CP-even Higgs bosons
each having a mass equal  to $\Lambda$, and each having a coupling
suppression relative to the SM of $R^2=1/(2+n)$. The excluded value of
$\Lambda$ is determined from the exclusion plot of  a Higgs boson with
a coupling equal to $R^2=1/(2+n)$. For example for $n=1,2,3$ we find
that $R^2=1/3,1/4,1/5$ corresponding to the LEP1 excluded values given
from Fig.4a of about $\Lambda>59,58,57.5$ GeV, respectively. We find it
remarkable that such strong limits on $\Lambda$ can be placed on models
with several singlets.

We emphasise that this argument is only approximate, and may become
unreliable when more realistic Higgs exclusion data as in Fig.4a are
used, rather than the simple analytic calculations of how large $R_i^2$
may become as a function of $m_{h_i}^2$ (as in Fig4b).

Finally note that if Higgs bosons are discovered then this does not
mean, in the context of this model, that $\Lambda$ is small. One may
have a large $\Lambda$ (as large as 146 GeV ) and still be lucky enough
to find that there is a visible CP-even Higgs boson corresponding to
some fortunate values of parameters. All we have shown is how the
non-observation of Higgs bosons enables a firm lower limit to be placed
on $\Lambda$, corresponding to a ``worst case'' situation. In general,
for a given $\Lambda$, the true situation will be easier than this,
leading to the possibility of Higgs discovery at LEP2 even for very
large $\Lambda$.

\subsection{$Z\to hA$}

In section 3 it was seen that the $Zh_iA_j$ couplings factorise into
a factor from the CP-odd matrix multiplied by a factor from the CP-even
matrix \cite{NMSSM} given by
\ba
R_{Zh_iA_1}&=&S_i\cos\gamma \cr
R_{Zh_iA_2}&=&-S_i\sin\gamma.
\ea
Using equation (\ref{XX}) we find
\ba
M^{\prime 2}_{22}&=& X_{12}^2m_{h_1}^2
                   + X_{22}^2m_{h_2}^2
                   + X_{32}^2m_{h_3}^2 \cr
               &=&S_1^2m_{h_1}^2+S_2^2m_{h_2}^2+S_3^2m_{h_3}^2
\label{basicA}
\ea
where we have used the fact that $S_i=X_{i2}$.
Again
\be
S_1^2+S_2^2+S_3^2=1
\label{unitarityA}
\ee
from the unitarity of $X$.
$M^{\prime 2}_{22}$ just corresponds to the less useful bound
in Eq.(\ref{bound1lessuseful}).
Eq.(\ref{basicA}) is of course virtually identical to Eq.(\ref{basic}),
and hence we may immmediately write down analagous
bounds which relate the masses of the lightest CP-even states
to their $ZhA$ couplings :
\ba
m_{h_1}^2 &\leq & M_{22}^{\prime 2}\cr
m_{h_2}^2 &\leq &\frac{M_{22}^{\prime 2}-S_1^2m_{h_1}^2}{1-S_1^2}\cr
m_{h_3}^2 &\leq &
\frac{M_{22}^{\prime 2}-(S_1^2+S_2^2)m_{h_1}^2}{1-S_1^2-S_2^2}
\label{newbounds}
\ea

It is trivial to prove that
\be
m_{A}^{\prime 2}=m_{A_1}^2\cos^2\gamma+m_{A_2}^2\sin^2\gamma
\label{basicA2}
\ee
which is similar to Eqs.(\ref{basic}) and (\ref{basicA}).
Eq.(\ref{basicA2}) implies the bound
\be
{m_{A_1}}^2\leq {m'_A}^2
\ee
so that ${m'_A}^2$ is just the upper bound on the
lightest CP-odd Higgs boson in this model.
When $A_1$ is weakly coupled ($\cos \gamma \approx 0$),
Eq.(\ref{basicA2}) implies that
${m_{A_2}}^2\approx {m'_A}^2$.
Thus if the lighter CP-odd
state is weakly coupled,
corresponding to its having a large singlet component,
then the heavier CP-odd state
(which essentially corresponds to the MSSM CP-odd state) sits at the bound,
in complete analogy with Eq.(\ref{equality3}).

We now discuss how LEP limits on the non-observation of $Z\rightarrow
hA$ may be used to exclude values of ${m'_A}^2$ and $M_{22}^{\prime 2}$
smaller than a certain amount. The following discussion obviously
parallels that of Section 5.3. As before, the  ``worst case'' for $Z\to
hA$ production will provide the best exclusion limit on ${m'_A}^2$ and
$M_{22}^{\prime 2}$. In order to determine this worst case we simply
repeat the algorithm in Section 5.3, but with the  $R_i$ couplings
replaced by the $S_i$ couplings in Eq.(\ref{newbounds}). However in the
present case we need to generalise this procedure to include the CP-odd
parameters $m_{A_1}$ and $\gamma$. Using the exclusion criterion of
requiring 50 $hA$ events for exclusion, it is readily seen that such a
procedure enables the maximum excluded values of  ${m'_A}^2$ and
$M_{22}^{\prime 2}$ to be obtained.

Similarly to the case of Section 5.3, we find the worst case to
correspond to the CP-odd parameters $m_{A_1}=m_{A_2}=m'_A$ and
$\gamma=\pi/4$,  and the CP-even parameters $S_1=S_2=S_3=1/\sqrt 3$ and
$m_{h_1}=m_{h_2}=m_{h_3}=M'_{22}$. This corresponds to
${R_{Zh_iA_j}}^2=1/6$ in each case. The excluded values of ${m'_A}^2$
and $M_{22}^{\prime 2}$ may be found by using the approximation of 50
$hA$ events with $h$ having a mass equal to $M'_{22}$, $A$ having a
mass equal to ${m'_A}$, and the coupling ${R_{ZhA}}^2=1/6$, similarly
to the $Z\to Zh$ case discussed above.

As before, the above rule of thumb is easily extended to the more
general case of $n$ singlets being added to the MSSM, {\em i.e.} the
(M+$n$)SSM. In such a model the worst case of $hA$ production will
correspond to $n+2$ CP-even Higgs bosons each having a mass equal  to
$\sqrt{M^{\prime 2}_{22}}$, and each having a coupling factor of
$S^2=1/(2+n)$, plus $n+1$ CP-odd Higgs bosons each having a mass equal
to $m'_A$, and each having a coupling factor of $P^2=1/(1+n)$. As
before, when the states are non-degenerate, there must be at least one
pair $h_i$ and $A_j$ which have a cross-section larger than this. For
example for $n=1,2,3$, the couplings are ${R_{ZhA}}^2=1/6,1/12,1/20$.

We expect this simple approximation to be rather more robust for the
$Z\to hA$ process than for $Z\to Zh$, because of the more rapid
fall-off of the cross-section with increasing masses for the $hA$
process than for $Zh$. Furthermore, the larger number of parameters
here make a more elaborate analysis, such as we shall later perform for
$Z\to Zh$, more difficult. Thus we shall always use this approximate
method when we present our results for $Z\to hA$.

We note that of course  ${m'_A}^2$ and $M_{22}^{\prime 2}$ are not
independent parameters. Using Eq.(\ref{bound1lessuseful}) the value of
$M_{22}^{\prime 2}$ may be related to ${m'_A}^2$ and, for a given
$\lambda$ and $\tan \beta$, $M_{22}^{\prime 2}$ may be eliminated.
Thus, for a fixed value of $\lambda$, we shall present exclusion plots
in the $m'_A-\tan \beta$ plane from the non-observation of
$Z\rightarrow hA$.

\subsection{Charged Higgs Detection}

As discussed above, the $Z\to hA$ and $Z\to Zh$ searches are less
powerful in extended models because the cross-section can be greatly
reduced. We now turn to the last feasible search at LEP2, that for
charged Higgs production. One feature of the model which was discussed
earlier was that the charged Higgs can be lighter than in the MSSM when
singlets are included; furthermore, its couplings cannot be suppressed
by singlet mixing. Hence the charged Higgs signal, which in the MSSM is
completely dominated by $Z\to hA$, is now far more important.

Charged Higgs discovery is complicated, since the rate is strongly
dependent not only on the Higgs mass, which must not be too close to
the $W$ or $Z$ mass to allow elimination of background, but also on its
branching fractions to $cs$ and $\tau\nu$ which are determined by
$\tan\beta$ \cite{charged}. In addition, any LEP2 discovery region may
be dominated by study of top quark decays at the TeVatron \cite{topch}.
For the purposes of this paper we shall adopt the rather optimistic
view that LEP2 will have sufficient luminosity so that the kinematic
limit may be approached.

\section{Exclusion Limits in the $m'_A$-$\tan\beta$ Plane}

In this section we shall present our results as excluded regions in the
$m'_A-\tan \beta$ plane, which is familiar from similar studies in the
MSSM, and should simplify the comparison of the (M+1)SSM with the MSSM.
Let us first briefly summarise the LEP2 search strategies we have so
far introduced in Section 5.

For the process $Z\rightarrow h_iA_j$ (discussed in section 5.4) we
have shown how LEP2 can be used to place exclusion limits on ${m'_A}^2$
and $M_{22}^{\prime 2}$. For a fixed value of $\lambda$ (and radiative
correction parameters) these limits may be interpreted as excluded
regions in the $m'_A-\tan \beta$ plane. We show both the worst case
mixing scenario and the simple scenario where there is no singlet
mixing. These are calculated using the simple approximation that
50 $Z\to hA$ events will be sufficient to ensure a discovery.

Similarly, for the processes $Z\rightarrow Zh_i$ (discussed in sections
5.1-5.3) we have shown how LEP2 can be used to place an exclusion limit
on the value of $\Lambda$. For a fixed value of $\lambda$ (and
radiative correction parameters) this can be interpreted as  an
exclusion limit on $\tan\beta$, independently of $m'_A$ ({\em i.e.} a
horizontal exclusion line in the $m'_A$-$\tan\beta$ plane.) However, as
we shall see, the resulting excluded region is not very large, so
we shall resort to a more powerful technique as discussed below.

This new technique exploits the fact that the upper 2$\times$2 block of
the CP-even mass squared matrix in Eq.(\ref{CPevenmass}) is completely
specified (for fixed $\lambda$) in the $m'_A$-$\tan\beta$ plane.
However, unlike the MSSM, the CP-even spectrum is not completely
specified since it depends on three remaining unknown real parameters
associated with singlet mixing ({\em i.e.} the dots in
Eq.(\ref{CPevenmass}).) Each choice of these unconstrained terms then
completely specifies the parameters $m_{h_i},R_i$, and we can test to
see if the resulting Higgs spectrum is excluded or allowed. We then
scan over all possible choices (which we parametrise as $m_{h_1}$,
$R_1$ and one other mixing angle) and if the resulting spectrum can
always be excluded by LEP2, then we conclude that this point in the
$m'_A$-$\tan\beta$ plane (for fixed $\lambda$) can be excluded.

Since we have presented two separate algorithms for generating
$Z\rightarrow Zh_i$ contours, it is worthwhile here mentioning some of
the advantages and disadvantages of the different techniques. The
method of scanning over the whole of parameter space is naturally
better than the algorithm based on the exclusion limit of $\Lambda$, in
the sense that it gives a better reach in the plane. This is clear from
the fact that we are considering a general 3$\times$3 mass matrix which
has the upper 2$\times$2 block given by our position in the
$m'_A-\tan\beta$ plane, while the bound algorithm in section 5.3 allows
any mass matrix which has the correct 11 component in a particular
basis (which is why the exclusion contours derived in this way do not
have any $m_A$ dependence). On the other hand, the scanning technique
is very CPU-intensive, and would rapidly become even more so if we
allowed extra singlet states. In addition, it is hard to understand
these results analytically, whereas the  simple arguments in section
5.3 based around Eq.(\ref{basic}) are much more straightforward. Hence
both techniques are worth considering.

\subsection{$\lambda=0$}

In figures 6-8 we consider the impact of mixing only, with $\lambda$
set equal to zero. This may not be as unreasonably optimistic as it
sounds, since recent GUT scale analyses \cite{ulrich,epic} have
concluded that very small $\lambda$ is preferred with universal soft
parameters.  In each case we show the charged Higgs kinematic limit as
a dot-dashed line. The two $Z\to hA$ exclusion contours are represented
by dashed lines and drawn with the assumption of worst case mixing
(left) and of no singlet mixing (right). The three $Z\to Zh$ exclusion
contours are represented by solid lines, and shown for the case of no
singlet mixing (uppermost line),  the full scan over parameter space
(middle line), and for the bound algorithm (lower horizontal line)
discussed in detail in section 5.3. The no singlet mixing lines with
$\lambda=0$ simply correspond to the MSSM.

Figure 6 shows the exclusion contours for $\sqrt s=175$GeV, and
integrated luminosity of 150pb${}^{-1}$ per experiment. Figure 7 is
identical but with $\sqrt s=192$GeV, while Figure 8 has $\sqrt
s=205$GeV and integrated luminosity of 300pb${}^{-1}$ per experiment.
All these figures have $m_t=175$GeV, degenerate squarks at 1TeV, and
$\alpha_3(M_Z)=0.12$.

In each of these figures it is clear that the charged Higgs  kinematic
limit discovery line (which, as mentioned above, is rather
over-optimistic as a discovery process, but is nevertheless indicative
of where this process will become important) is completely superseded
by the $Z\to hA$ line, although to a rather less extreme degree than in
the MSSM. The inclusion of singlets reduces the excluded region for
both $Z\to hA$ and $Z\to Zh$ quite substantially relative to the MSSM,
but still leaves reasonably large areas covered. Simply using the bound
algorithm however gives rather poorer reach.

\subsection{$\lambda>0$}

To show how large $\lambda$ affects our results, in Figure 9 we show a
figure with $\sqrt s=192$GeV, and integrated luminosity of
150pb${}^{-1}$ per experiment, but with $\lambda=0.5$. Here the charged
line is becoming quite competitive with the $Z\to hA$ line, and so a
charged Higgs search may well be the most practical one at LEP2. The
$Z\to hA$ contours are hardly changed from those in the $\lambda=0$
case except for very small $\tan\beta$.

The most significant impact is however on the $Z\to Zh$ lines. Here
even the no mixing scenario has very little reach in the
$m_A-\tan\beta$ plane because, as is clear from Eq.(\ref{bound1}),
small $\tan\beta$ no longer implies a light CP-even state. With singlet
mixing, no part of the plane can be covered at all.

\section{Conclusion}

The addition of extra singlets to the MSSM greatly complicates the
model and renders far harder the main experimental searches.  However
it is possible to exclude this model by placing an experimental lower
bound $\Lambda_{min}$ on the value of the parameter $\Lambda$ which is
just the upper bound on the lightest CP-even Higgs boson mass. In the
limit that the lightest CP-even Higgs boson is very weakly coupled,
this bound applies to the second lightest CP-even Higggs boson. Thus
for a fixed value of $\Lambda$ the entire CP-even Higgs boson spectrum
is constrained, resulting in experimentally excluded values of
$\Lambda$ associated  with the ``worst case'' scenarios discussed in
section 5.3.

We have seen that LEP1 already finds $\Lambda>\Lambda_{min}=59$ GeV,
and LEP2 will set limits of $\Lambda_{min}\approx 81,93,105$ GeV for
three different  levels of operation. The theoretical upper bound on
$\Lambda$ is $\Lambda_{max} \approx 146$ GeV, depending on the details
of the squark spectrum and on the top mass. Once $\Lambda_{min}$
becomes greater than or equal to $\Lambda_{max}$ then the model will be
excluded. We have generalised this procedure to the case of an
arbitrary number of extra singlets. Thus our first conclusion is that
it will be possible to exclude a version of the MSSM containing
additional singlets.

The effects of the additional singlet can be thought of as reducing the
Higgs sector of the model to the MSSM with two additional complicating
factors : extra singlet states which can mix in an arbitrary way with
the usual neutral Higgs states altering the masses and diluting the
couplings of the mass eigenstates; and an extra Higgs sector quartic
coupling  $\lambda$ which changes the mass matrices even in the absence
of singlet mixing.

We have systematically studied the case  of singlet mixing with
$\lambda=0$, with the primary conclusion that, while the inclusion of a
singlet can significantly complicate matters for searches at LEP2, it
is still possible to cover a significant amount of the $m_A'-\tan\beta$
plane using the usual $Z\to Zh$ and $Z\to hA$ searches. The main
difference in strategy between the searches in the MSSM and in models
with singlets is that more luminosity is needed to cover the same area
of the plane, since states can be more weakly coupled than in the MSSM.
A second difference is that, because singlets cannot mix with charged
states, the charged Higgs signal cannot be degraded and so is more
important than in the MSSM where it is completely dominated by $Z\to
hA$.

The effect on non-zero $\lambda$ is more troublesome. For large
$\lambda$ the mass bounds in the MSSM are markedly increased,
particularly at small $\tan\beta$ which is of course the region where
LEP2 normally has the best reach. This largely wrecks the prospect of
excluding large parts of the $m_A'-\tan\beta$ plane through the process
$Z\to Zh$, and has some impact on $Z\to hA$. However, we note that
large $\lambda$ also reduces the charged Higgs mass substantially,
making its discovery easier.

We conclude on a positive note by pointing out that for much of this
paper we have assumed worst case mixing scenarios, and considered how
it might be possible to rule out regions of parameter space in a
consistent way for arbitrary parameters from the singlet sector. For
the MSSM, specifying a point in the $m_A-\tan\beta$ plane completely
specifies all the masses and mixings, and so points outside the
discovery contours cannot be discovered; however with singlets it is
possible that, if we are lucky enough, {\em any} point in the
$m_A'-\tan\beta$ plane could lead to a discovery, since it is possible
to construct singlet mixing parameters which will reduce the mass of a
heavy CP-even state while leaving its coupling reasonably large. Hence
we can argue that despite the rather negative impact of the mixing on
the LEP2 exclusion contours, a SUSY Higgs discovery may be no less
likely in a model with singlets than in the MSSM.

\begin{center}

{\bf Acknowledgements}

\end{center}
We would like to thank the LEP2 Higgs working group and particularly
M.~Carena, U.~Ellwanger, H.~Haber, C.~Wagner, and P.~Zerwas for many
stimulating discussions and for encouraging us finally to do some work
on this topic. We are also very grateful to P.~Janot and A.~Sopczak for
sharing their data and expertise.

\newpage
\section{Figure Captions}

\noindent
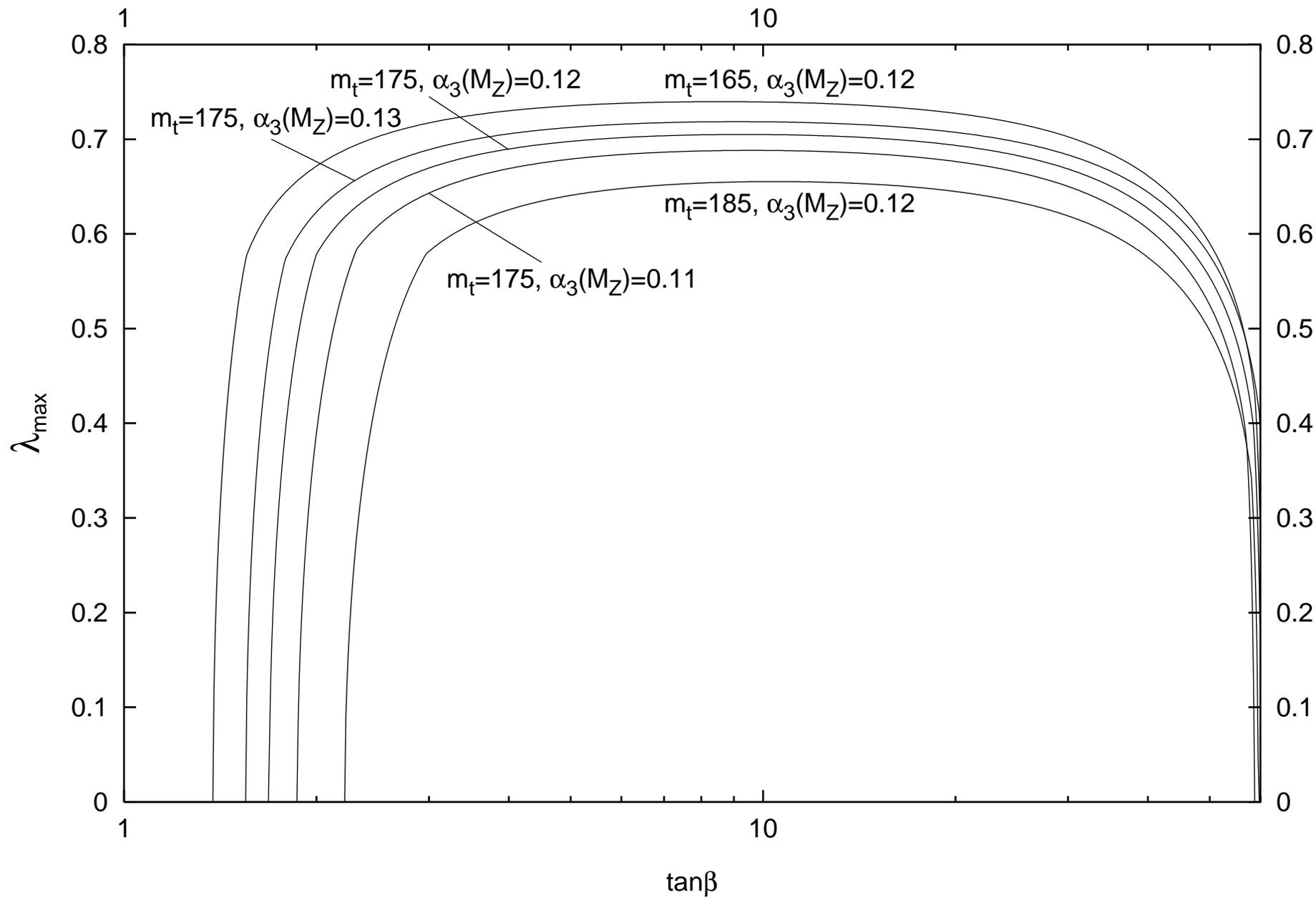
{\bf Figure 1:}
Maximum value of $\lambda(m_t)$ as a function of $\tan\beta$, shown for
various values of $m_t$ and $\alpha_3(M_Z)$ and derived from requiring
perturbativity of all Yukawa couplings up to a scale $10^{16}$GeV.

\noindent
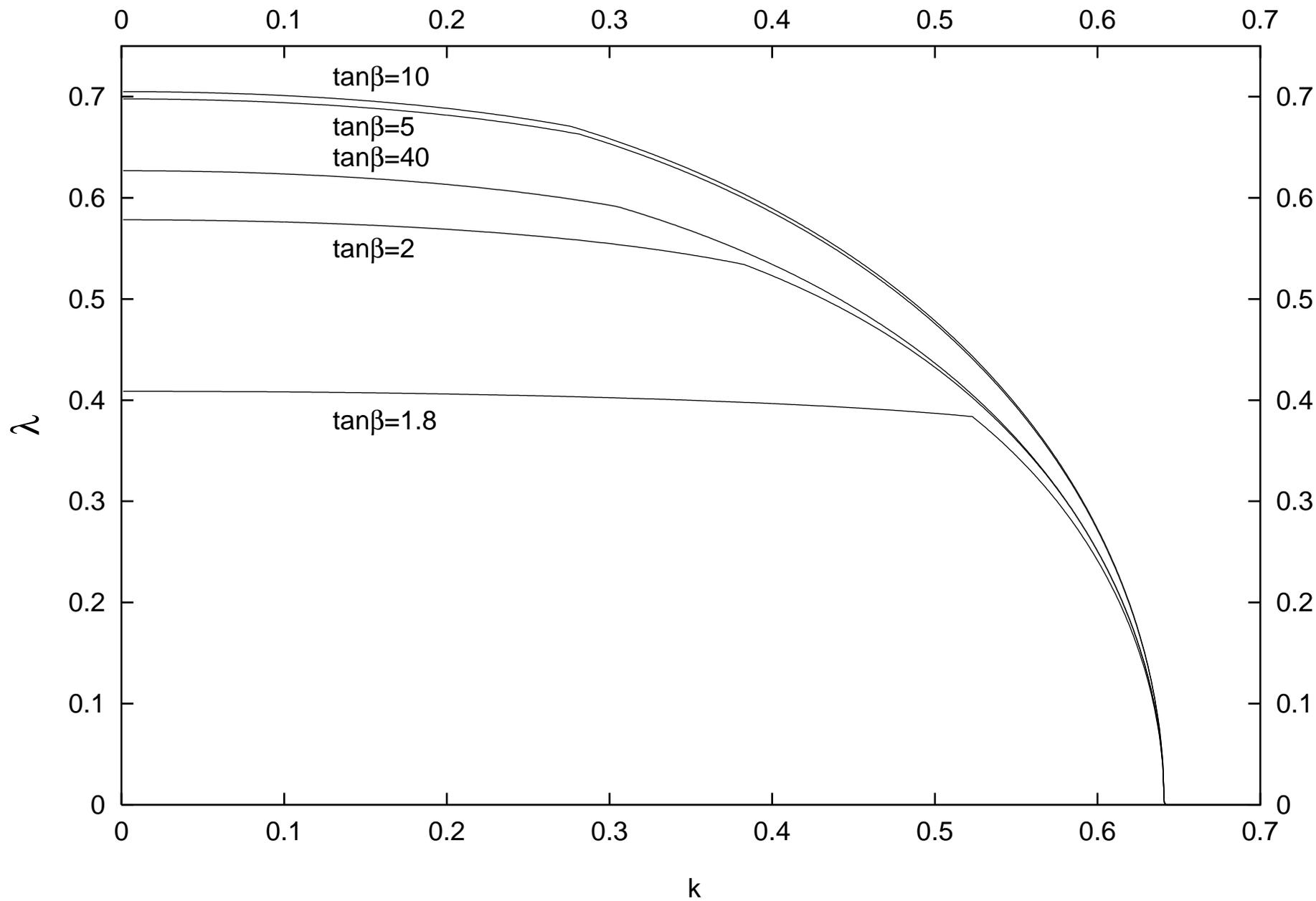
{\bf Figure 2:}
Maximum value of $\lambda(m_t)$ as a function of $k(m_t)$, shown for
various values of $\tan\beta$, with $m_t=175$GeV and
$\alpha_3(M_Z)=0.12$ and derived from requiring perturbativity of all
Yukawa couplings up to a scale $10^{16}$GeV.

\noindent
{\bf Figure 3a:}
$\Lambda_{max}$, the bound on the lightest Higgs in the MSSM with
(upper solid line) and without (lower dashed line) added singlets,
shown as a function of $\tan\beta$ with $M_{susy}=1$TeV, $m_t=175$GeV,
and $\lambda$ and $A_t$ set so as to maximise the bound.

\noindent
{\bf Figure 3b:}
$\Lambda_{max}$, the bound on the lightest Higgs in the MSSM with
(upper solid line) and without (lower dashed line) added singlets,
shown as a function of $M_{susy}$ with  $m_t=175$GeV, and $\lambda$,
$A_t$ and $\tan\beta$ set so as to maximise the bound.

\noindent
{\bf Figure 3c:}
$\Lambda_{max}$, the bound on the lightest Higgs in the MSSM with
(upper solid line) and without (lower dashed line) added singlets,
shown as a function of $m_t$ with  $M_{susy}=175$GeV, and $\lambda$,
$A_t$ and $\tan\beta$ set so as to maximise the bound.

\noindent
{\bf Figure 4a:}
$R^2$ versus $m_h$ for three different sets of machine parameters at
LEP2, using a full analysis of the experimental data. For
$m_h\stackrel{<}{\sim}$50GeV the LEP1 bounds are tighter and so we use
those instead. These are 95\% confidence level exclusion contours.
Points above the line correspond to states which could be seen, those
below to those which could not be seen. The three lines are for
$\sqrt{s}=175$GeV, $\int{\cal{L}}=150$pb${}^{-1}$ per experiment;
$\sqrt{s}=192$GeV, $\int{\cal{L}}=150$pb${}^{-1}$ per experiment;
$\sqrt{s}=205$GeV, $\int{\cal{L}}=300$pb${}^{-1}$ per experiment;
reading from left to right.

\noindent
{\bf Figure 4b:}
$R^2$ versus $m_h$ for three different sets of machine parameters at
LEP2, using the simple approximation that 50 CP-even events are
sufficient for detection of a state. Points above the line correspond
to states which could be seen, those below to those which could not be
seen. The three lines
$\sqrt{s}=175$GeV, $\int{\cal{L}}=150$pb${}^{-1}$ per experiment;
$\sqrt{s}=192$GeV, $\int{\cal{L}}=150$pb${}^{-1}$ per experiment;
$\sqrt{s}=205$GeV, $\int{\cal{L}}=300$pb${}^{-1}$ per experiment;
reading from left to right.

\noindent
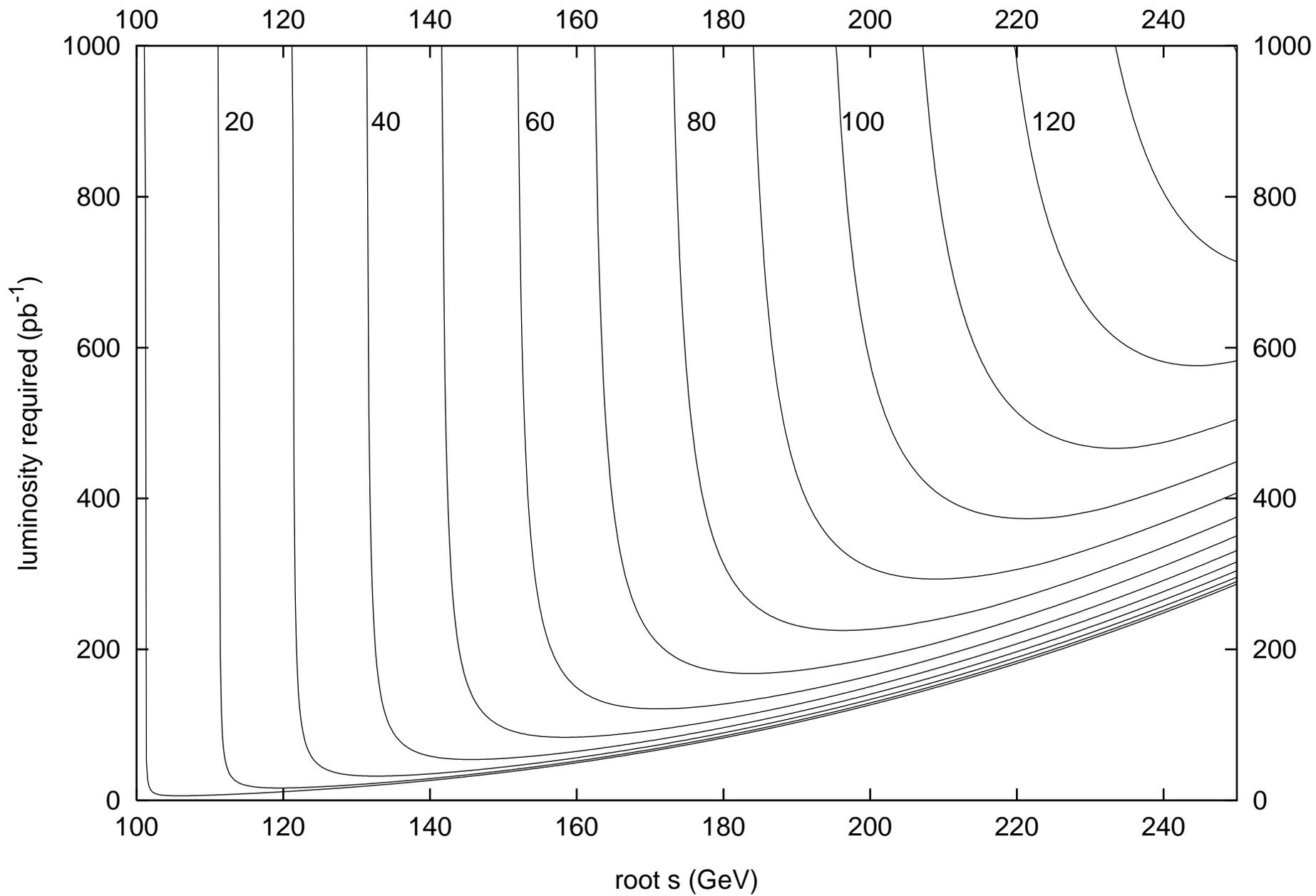
{\bf Figure 5:}
Contours of values of $\Lambda_{min}$, the value of $\Lambda$ which can
be excluded, in the  $\sqrt s-$luminosity plane, labelled in GeV. This
is total luminosity, not luminosity per experiment.

\noindent
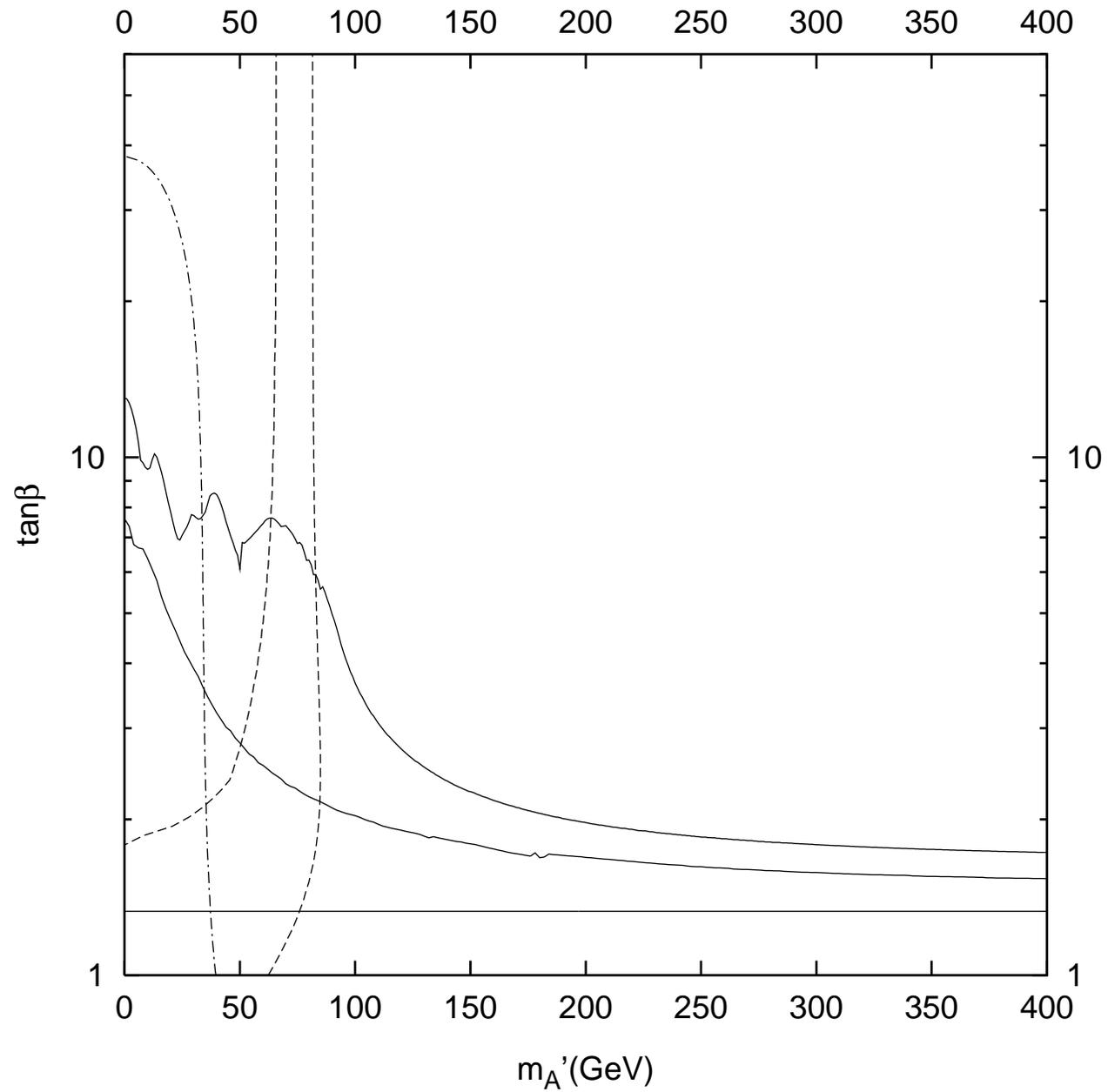
{\bf Figure 6:}
$m_A'-\tan\beta$ plane with $\sqrt s$=175GeV, $\int{\cal L}=150{\rm
pb}^{-1}$ per experiment, $\lambda=0$, $m_t=175$GeV, and squarks
degenerate at 1TeV. The dot-dashed line is the kinematic limit for
charged Higgs discovery. The dashed lines represent $Z\to hA$ contours
with no singlet mixing (right, corresponding to the MSSM) and worst
case mixing (left). The three solid lines show $Z\to Zh$ contours,
derived by assuming no singlet mixing (upper, corresponding to the
MSSM), the full scan approach (middle) and the exclusion
limit on $\Lambda$ (lower).

\noindent
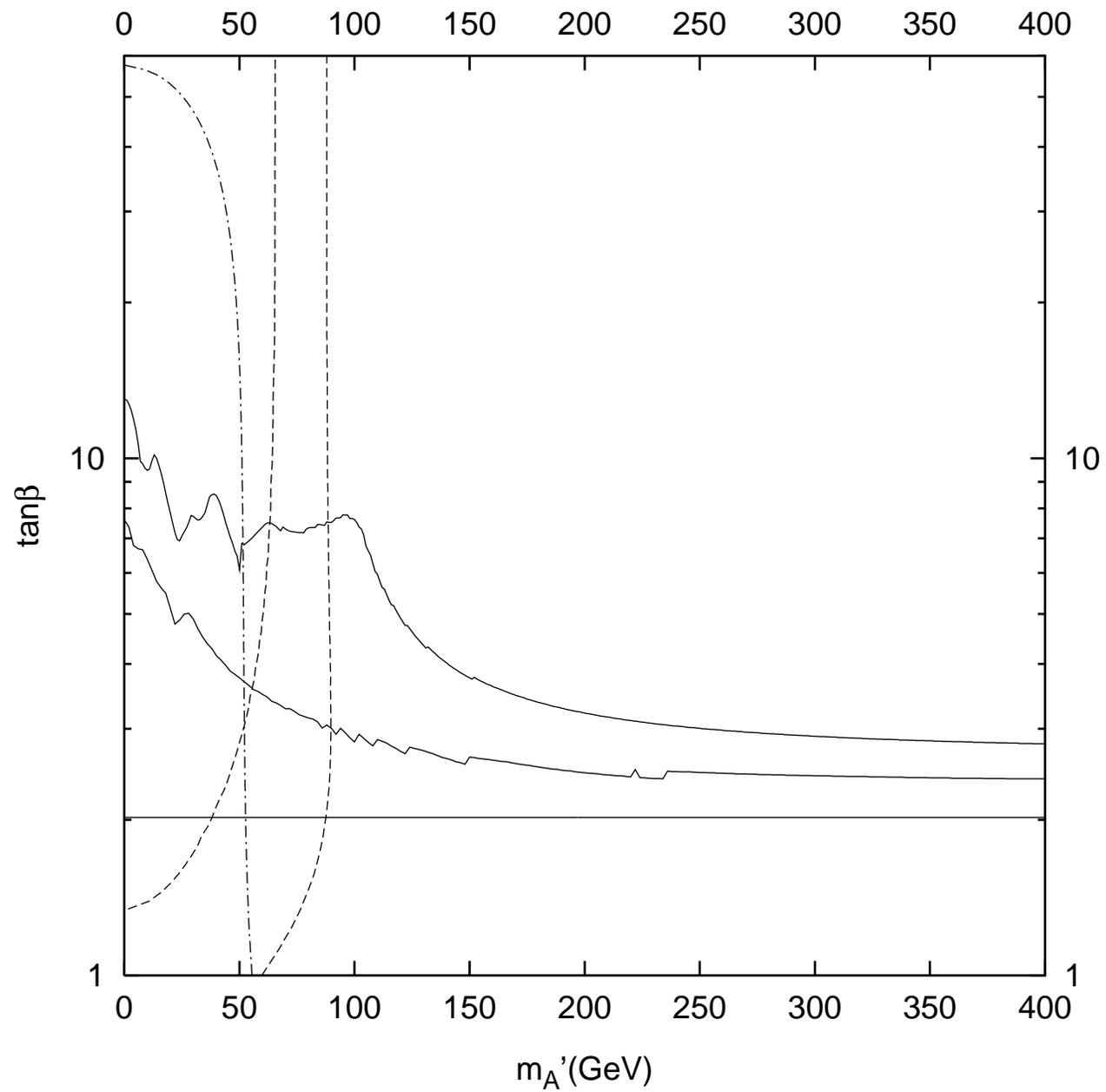
{\bf Figure 7:}
$m_A'-\tan\beta$ plane with $\sqrt s$=192GeV, $\int{\cal L}=150{\rm
pb}^{-1}$ per experiment, $\lambda=0$, $m_t=175$GeV, and squarks
degenerate at 1TeV. The dot-dashed line is the kinematic limit for
charged Higgs discovery. The dashed lines represent $Z\to hA$ contours
with no singlet mixing (right, corresponding to the MSSM) and worst
case mixing (left). The three solid lines show $Z\to Zh$ contours,
derived by assuming no singlet mixing (upper, corresponding to the
MSSM), the full scan approach (middle) and the exclusion
limit on $\Lambda$ (lower).

\noindent
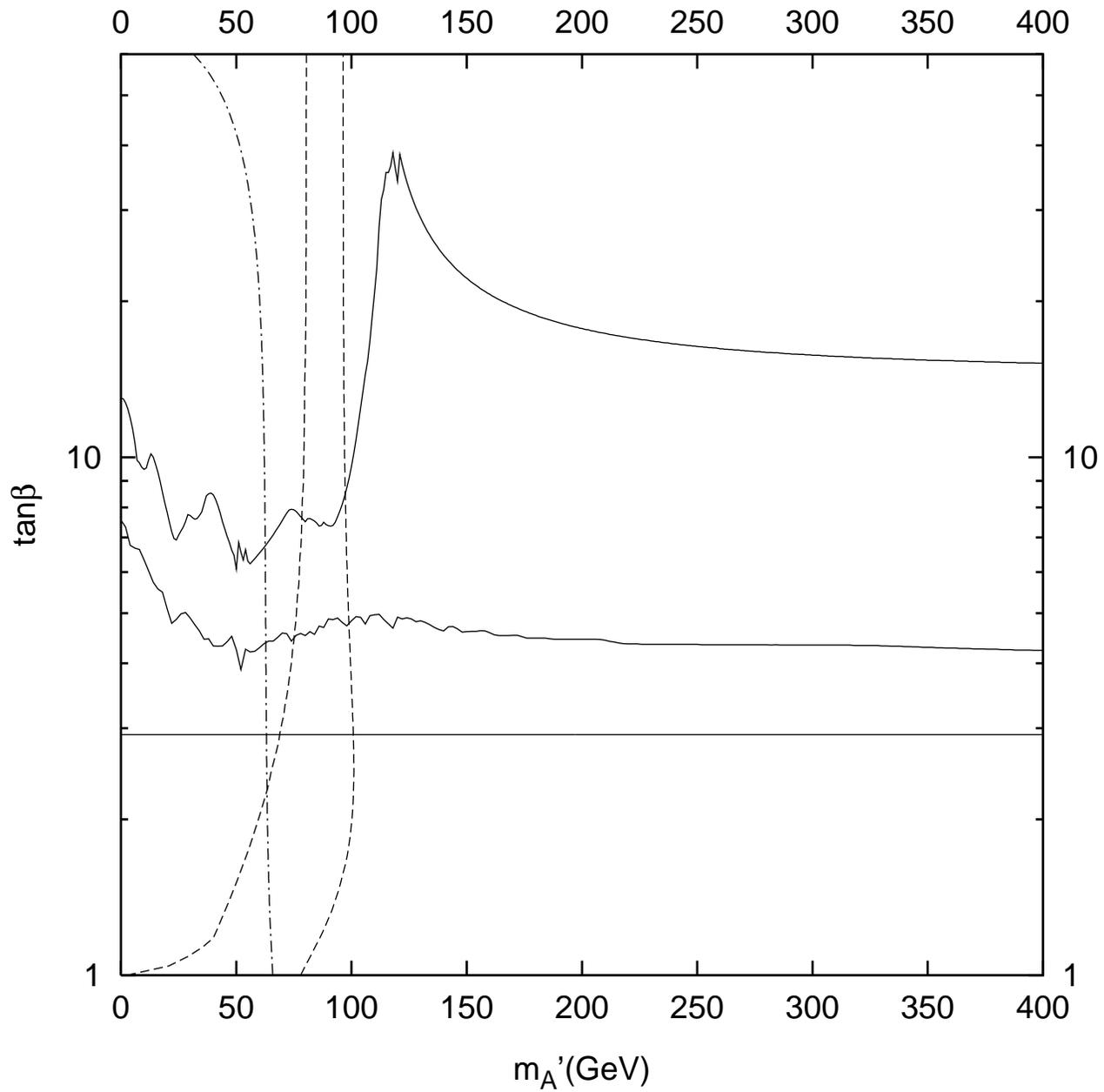
{\bf Figure 8:}
$m_A'-\tan\beta$ plane with $\sqrt s$=205GeV, $\int{\cal L}=300{\rm
pb}^{-1}$ per experiment, $\lambda=0$, $m_t=175$GeV, and squarks
degenerate at 1TeV. The dot-dashed line is the kinematic limit for
charged Higgs discovery. The dashed lines represent $Z\to hA$ contours
with no singlet mixing (right, corresponding to the MSSM) and worst
case mixing (left). The three solid lines show $Z\to Zh$ contours,
derived by assuming no singlet mixing (upper, corresponding to the
MSSM), the full scan approach (middle) and the exclusion
limit on $\Lambda$ (lower).

\noindent
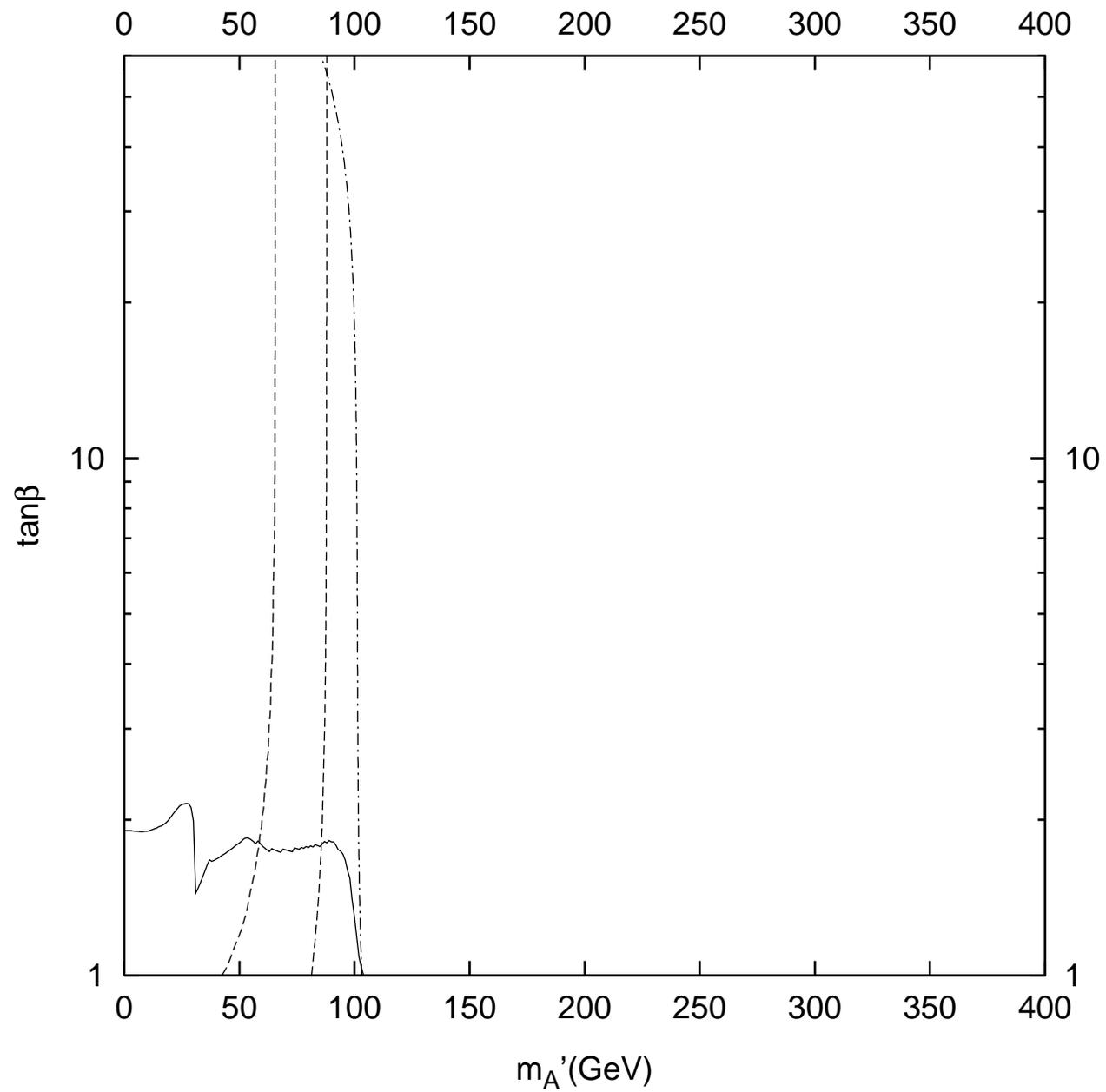
{\bf Figure 9:}
$m_A'-\tan\beta$ plane with $\sqrt s$=192GeV, $\int{\cal L}=150{\rm
pb}^{-1}$ per experiment, $\lambda=0.5$, $m_t=175$GeV, and squarks
degenerate at 1TeV. The dot-dashed line is the kinematic limit for
charged Higgs discovery. The dashed lines represent $Z\to hA$ contours
with no singlet mixing (right) and worst case mixing (left). The solid
line shows the $Z\to Zh$ contours derived by assuming no mixing. The
full scan approach and using the limit on $\Lambda$ do not allow any
part of the plane to be excluded.

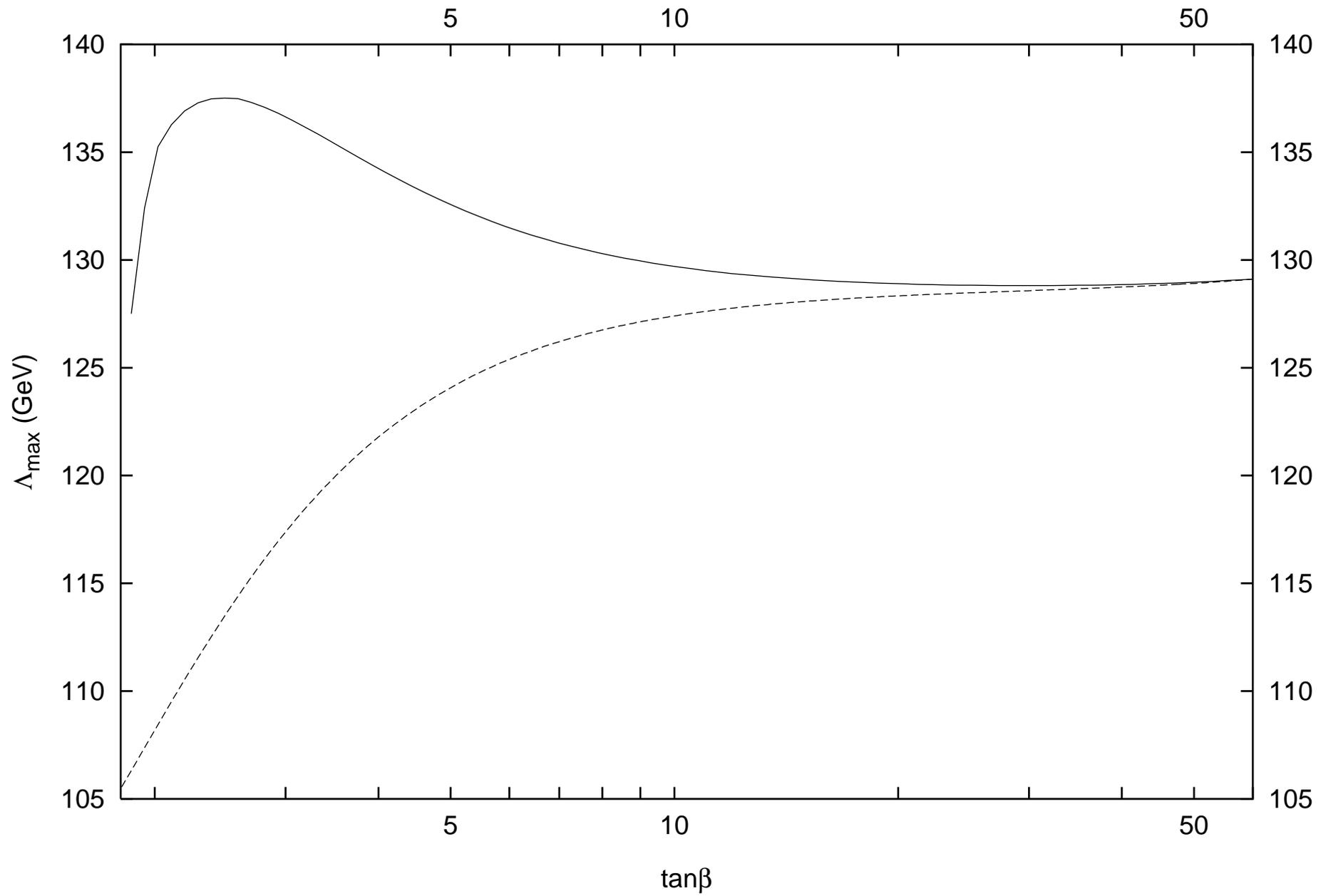
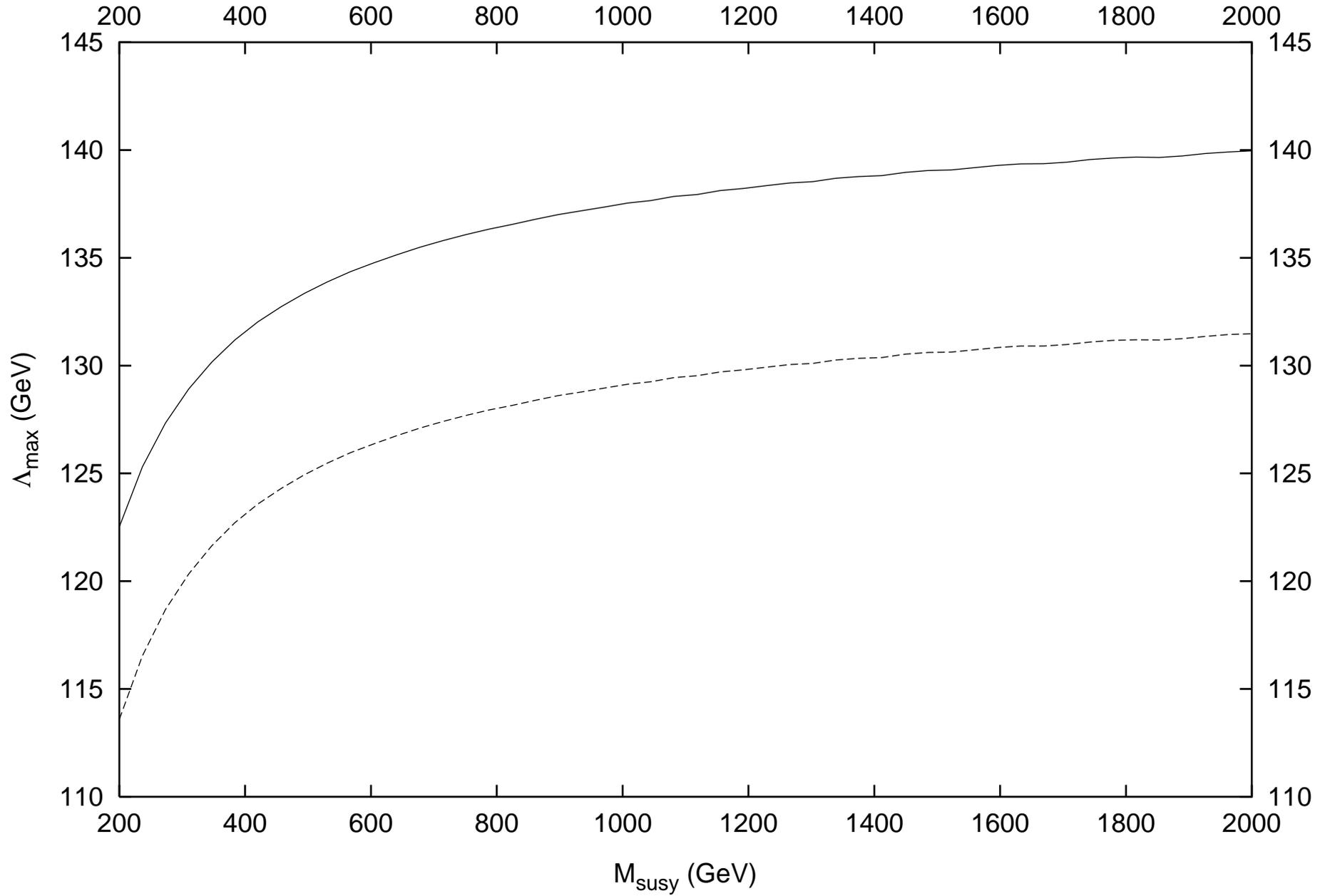
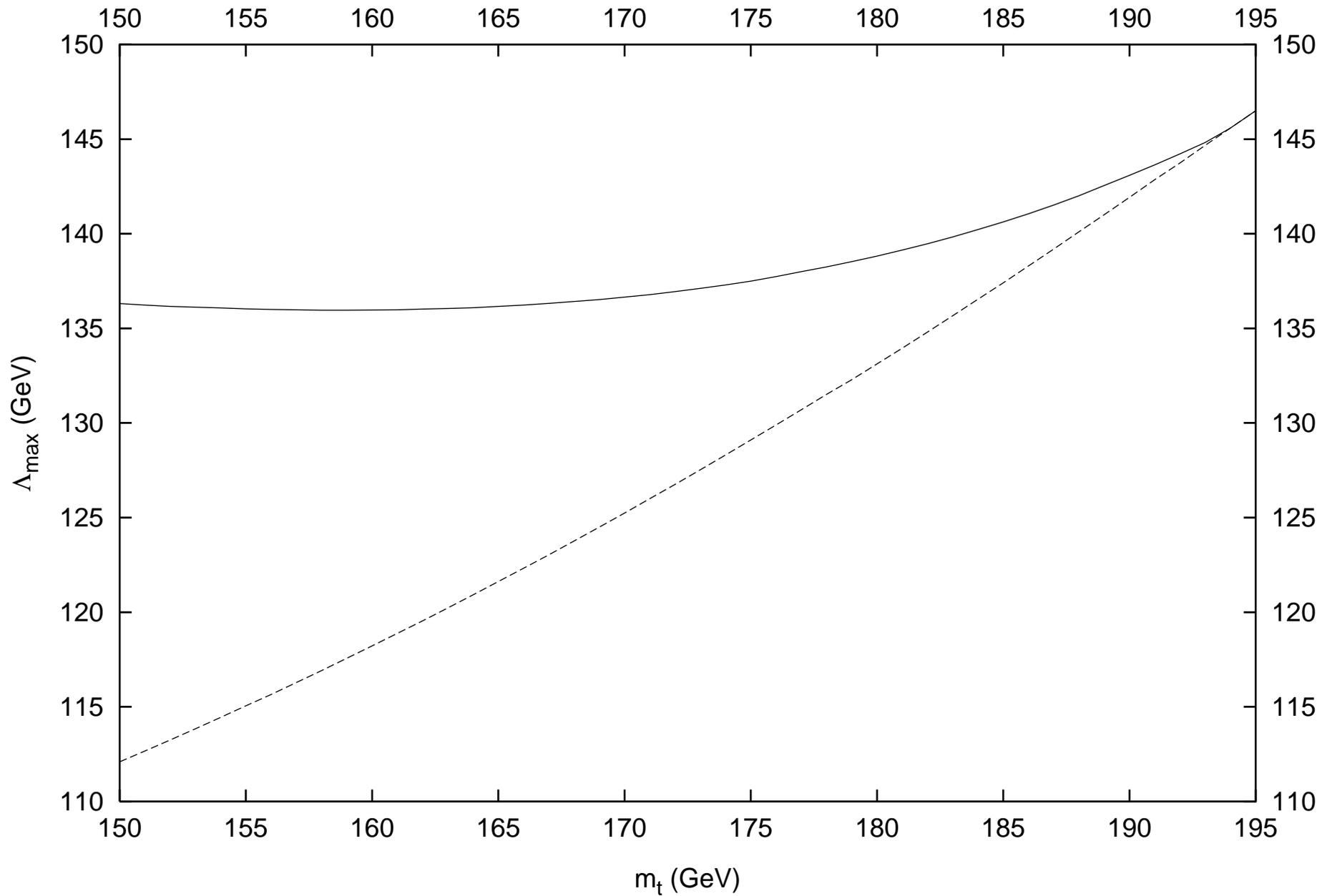
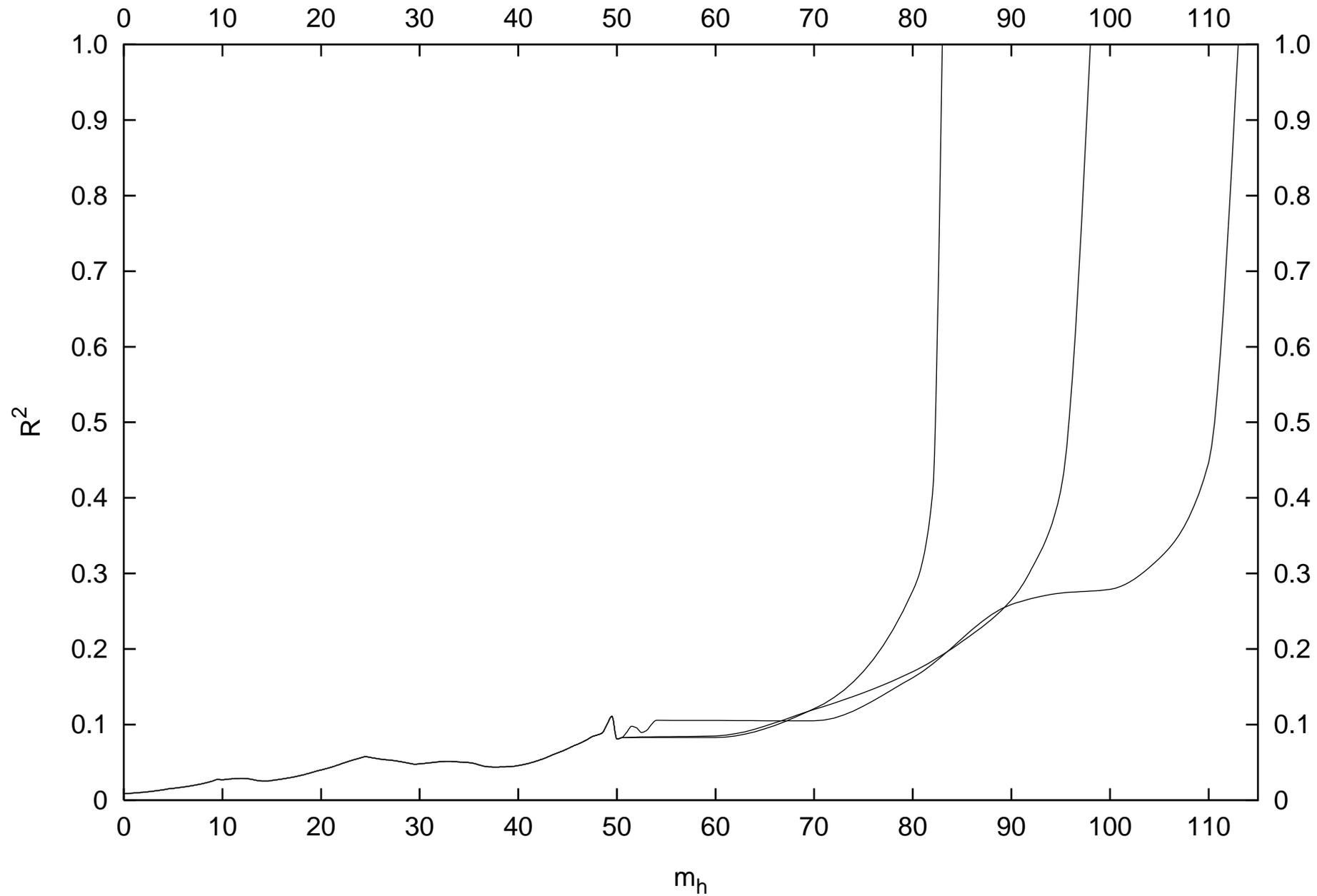
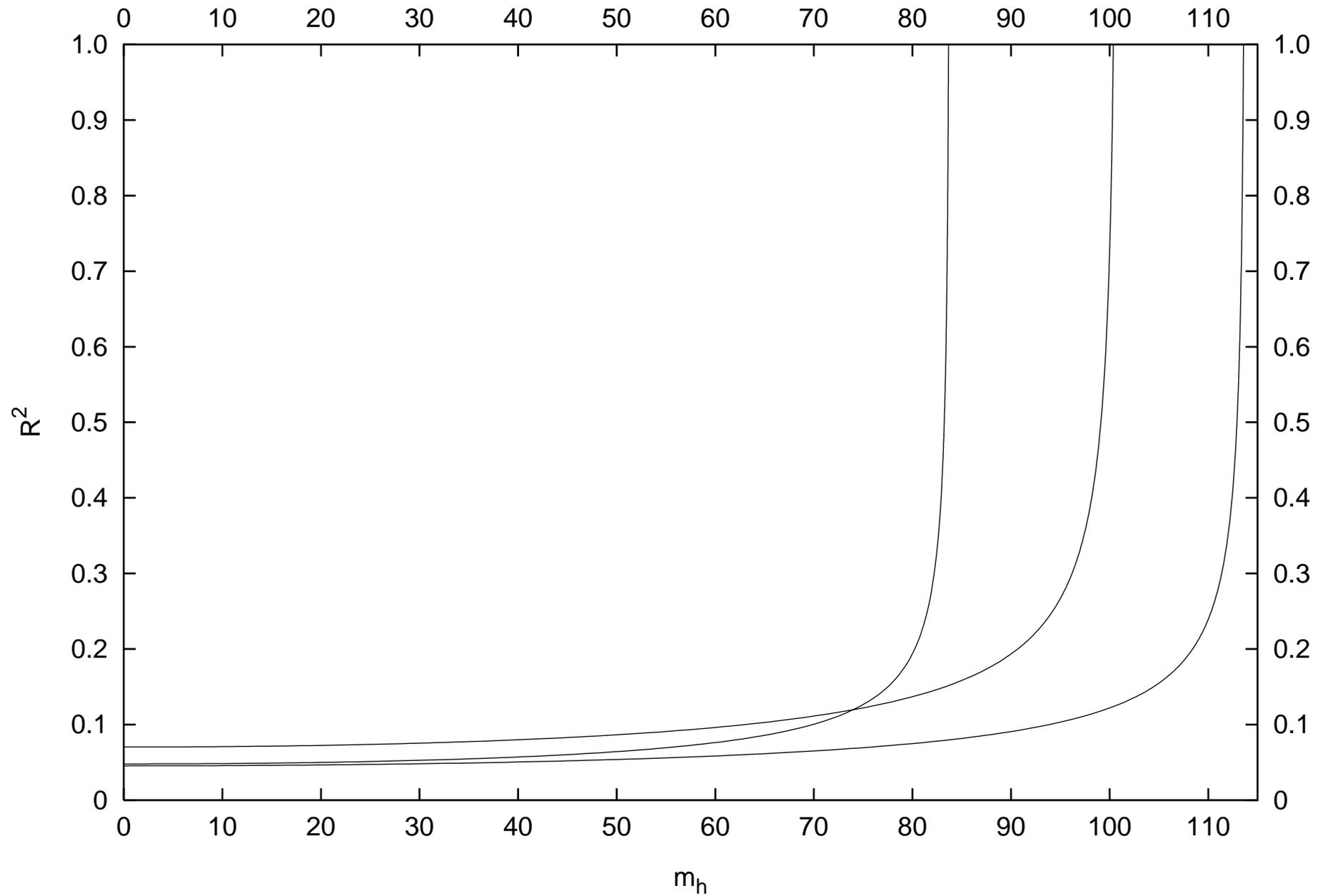
\end{document}